\documentstyle[epsf,onecolumn]{mn}

\newif\ifAMStwofonts

\input{psfig}
\def\ha{\rm H\alpha}
\def\Ha{H$\alpha$}

\def\hii{\hbox{H{\thinspace}II}}
\def\hi{\hbox{H{\thinspace}I}}
\def\etal{{\it et\thinspace al.}~}
\def\eg{{\it e.g.,}~}
\def\ergs{{\rm\,erg\,s^{-1}}}
\def\kms{{\rm\,km\ s^{-1}}}
\def\cc{{\rm\,cm^{-3}}}

\def\msol{\rm\,M_\odot}
\def\spose#1{\hbox to 0pt{#1\hss}}
\def\Dt{\spose{\raise 1.5ex\hbox{\hskip3pt$\mathchar"201$}}}    
\def\dt{\spose{\raise 1.0ex\hbox{\hskip2pt$\mathchar"201$}}}    

\def\Lha{\cal L}
\def\HII{\tiny\hbox{H{\thinspace}II}}
\def\HI{\tiny\hbox{H{\thinspace}I}}

\ifoldfss
  \ifCUPmtlplainloaded \else
    \NewTextAlphabet{textbfit} {cmbxti10} {}
    \NewTextAlphabet{textbfss} {cmssbx10} {}
    \NewMathAlphabet{mathbfit} {cmbxti10} {} 
    \NewMathAlphabet{mathbfss} {cmssbx10} {} 
  \fi
  \ifAMStwofonts
    \ifCUPmtlplainloaded \else
      \NewSymbolFont{upmath} {eurm10}
      \NewSymbolFont{AMSa} {msam10}
      \NewMathSymbol{\upi}     {0}{upmath}{19}
      \NewMathSymbol{\umu}     {0}{upmath}{16}
      \NewMathSymbol{\upartial}{0}{upmath}{40}
      \NewMathSymbol{\leqslant}{3}{AMSa}{36}
      \NewMathSymbol{\geqslant}{3}{AMSa}{3E}

      \let\leq=\leqslant 
      \let\geq=\geqslant 
    \fi
  \fi
\fi 

\ifnfssone
  \newmathalphabet{\mathit}
  \addtoversion{normal}{\mathit}{cmr}{m}{it}
  \addtoversion{bold}{\mathit}{cmr}{bx}{it}
  \newmathalphabet{\mathbfit} 
  \addtoversion{normal}{\mathbfit}{cmr}{bx}{it}
  \addtoversion{bold}{\mathbfit}{cmr}{bx}{it}
  \newmathalphabet{\mathbfss} 
  \addtoversion{normal}{\mathbfss}{cmss}{bx}{n}
  \addtoversion{bold}{\mathbfss}{cmss}{bx}{n}
  \ifAMStwofonts
    \ifCUPmtlplainloaded \else
      \UseAMStwoboldmath
      \makeatletter
      \new@mathgroup\upmath@group
      \define@mathgroup\mv@normal\upmath@group{eur}{m}{n}
      \define@mathgroup\mv@bold\upmath@group{eur}{b}{n}
      \edef\UPM{\hexnumber\upmath@group}
      \new@mathgroup\amsa@group
      \define@mathgroup\mv@normal\amsa@group{msa}{m}{n}
      \define@mathgroup\mv@bold\amsa@group{msa}{m}{n}
      \edef\AMSa{\hexnumber\amsa@group}
      \makeatother
      \mathchardef\upi="0\UPM19
      \mathchardef\umu="0\UPM16
      \mathchardef\upartial="0\UPM40
      \mathchardef\leqslant="3\AMSa36
      \mathchardef\geqslant="3\AMSa3E

      \let\leq=\leqslant 
      \let\geq=\geqslant 
    \fi
  \fi
\fi 

\ifnfsstwo
  \DeclareMathAlphabet{\mathbfit}{OT1}{cmr}{bx}{it}
  \SetMathAlphabet\mathbfit{bold}{OT1}{cmr}{bx}{it}
  \DeclareMathAlphabet{\mathbfss}{OT1}{cmss}{bx}{n}
  \SetMathAlphabet\mathbfss{bold}{OT1}{cmss}{bx}{n}
  \ifAMStwofonts
    \ifCUPmtlplainloaded \else
      \DeclareSymbolFont{UPM}{U}{eur}{m}{n}
      \SetSymbolFont{UPM}{bold}{U}{eur}{b}{n}
      \DeclareSymbolFont{AMSa}{U}{msa}{m}{n}
      \DeclareMathSymbol{\upi}{0}{UPM}{"19}
      \DeclareMathSymbol{\umu}{0}{UPM}{"16}
      \DeclareMathSymbol{\upartial}{0}{UPM}{"40}
      \DeclareMathSymbol{\leqslant}{3}{AMSa}{"36}
      \DeclareMathSymbol{\geqslant}{3}{AMSa}{"3E}

      \let\leq=\leqslant 
      \let\geq=\geqslant 
    \fi
  \fi
\fi 

\ifCUPmtlplainloaded \else
  \ifAMStwofonts \else 
    \def\upi{\pi}
    \def\umu{\mu}
    \def\upartial{\partial}
  \fi
\fi

\title[Superbubble Size Distribution]
	{The Superbubble Size Distribution in the \\ Interstellar Medium 
	of Galaxies}

\author[M. S. Oey and C. J. Clarke]{M. S. Oey\thanks
	{Email:  oey@ast.cam.ac.uk (MSO); cclarke@ast.cam.ac.uk (CJC).}
	and C. J. Clarke$^\star$ \\
	Institute of Astronomy, University of Cambridge, Madingley Road,
	Cambridge   CB3 0HA}

\date{Accepted 1997 March 03.
      Received 1996 December 10;
      in original form 1996 December 10}

\pagerange{\pageref{firstpage}--\pageref{lastpage}}
\pubyear{1997}

\begin{document}

\maketitle

\label{firstpage}

\begin{abstract}
We use the standard, adiabatic shell evolution to predict the
differential size distribution $N(R)$ for populations of OB
superbubbles in a uniform ISM.  Assuming that shell growth stalls upon
pressure equilibrium with the ambient ISM, we derive $N(R)$ for simple
cases of superbubble creation rate and mechanical luminosity function (MLF).
For constant creation and an MLF $\phi(L)\propto L^{-\beta}$,
we find that $N(R)\propto R^{1-2\beta}$ for $R<R_e$, and $N(R)\propto
R^{4-5\beta}$ for $R>R_e$, where the characteristic radius $R_e\sim
1300$~pc for typical ISM parameters.  For
$R<R_e,\ N(R)$ is dominated by stalled objects, while for $R>R_e$ it
is dominated by growing objects.  The relation $N(R)\propto R^{1-2\beta}$
appears to be quite robust, and also results from 
momentum-conserving shell evolution.  We predict a peak in
$N(R)$ corresponding to individual SNRs, and suggest that the contribution 
of Type~Ia SNRs should be apparent in the observed form of $N(R)$.  We present
expressions for the porosity parameters, $Q_{\rm 2D}$ and $Q_{\rm 3D}$,
derived from our analysis.  $Q_{\rm 2D}$ is dominated by the largest
superbubbles for $\beta < 2$ and individual SNRs for $\beta> 2$,
whereas $Q_{\rm 3D}$ is normally dominated by the few largest shells.

We examine evolutionary effects on the \hii\ region luminosity
function (\hii\ LF), in order to estimate $\beta$.
We find that for a nebular luminosity fading with time $t$, 
${\Lha}\propto t^{-\eta}$, there is a minimum observed slope $a_{\rm
min}$ for the \hii\ LFs.
Empirical measurements all show $a>a_{\rm min}$, therefore
implying that usually we may take $\beta = a$.
We also find that if nebular luminosity is instantaneously
extinguished at some given age, rather than continuously fading, no
$a_{\rm min}$ will be observed.

Comparison with the largely complete \hi\ hole catalog for the SMC
shows surprising agreement in the predicted and observed slope of $N(R)$.
This suggests that no other fundamental process is
needed to explain the size distribution of shells in the SMC.
Further comparison with largely incomplete \hi\ data for M31, M33,
and Holmberg~II also shows agreement in the slopes, but perhaps
hinting at systematic differences between spiral and Im galaxies.
We estimate porosities that are substantially $<1$ for all
of the galaxies except Holmberg~II, for which we obtain values
$\ga 1$.  Most of these galaxies therefore may not be strongly
dominated by a hot interstellar component.  However, porosity results
for the Galaxy remain inconclusive with the available data. 

\end{abstract}

\begin{keywords}
ISM: bubbles --- ISM: general --- \hii\ regions --- ISM:
structure --- supernova remnants --- galaxies: ISM --- galaxies:
individual: SMC
\end{keywords}

\section{Introduction}

The evolution of superbubbles created by the stellar winds and
supernovae of OB associations is one of the primary processes that
determines the structure and energetics of the interstellar medium
(ISM).  All components of the diffuse interstellar medium, including 
warm and cold \hi\, as well as the warm ionized medium (WIM), are
thought to have large fractions of their volume consisting largely of
superbubble walls (\eg Heiles 1984; Kennicutt {\etal}1995).  The
restructuring of cool gas will in turn 
influence the collapse of molecular clouds and attendant star
formation (\eg McCray \& Kafatos 1987).  Likewise, the hot ionized
medium (HIM) is believed to originate within these superbubbles and
Type Ia supernova remnants (SNRs), 
hence the evolution of these structures determines whether and how
this gas is released into the general ISM.  
Models for the ISM are strongly dependent on whether the coronal gas
is the pervasive, dominant component, as envisioned by \eg McKee \&
Ostriker (1977), or whether it plays a lesser role as
favored by \eg Slavin \& Cox (1993), returning to a
paradigm closer to the two-phase ISM of Field, Goldsmith, \&
Habing (1969).

The standard model for understanding the evolution of superbubbles and
other shell structures is that of an adiabatic, pressure-driven
bubble, with continuous wind energy injection (\eg Weaver
{\etal}1977; Dyson 1977; Pikel'ner 1968).  It has generally been
assumed that in later stages,  
successive supernovae (SNe) can be treated as an approximate
continuous energy injection, and are able to power the growth of
supergiant shells (\eg Mac Low \& McCray 1988).
Such supergiant shells, having radii of hundreds of pc, could then
blow out of the galactic disk, releasing hot gas into the 
halo, possibly in a galactic fountain cycle.

However, recent evidence suggests that superbubble evolution is often
not as simple as described by the standard model.
Many superbubbles exhibit expansion
velocities that are too high to be consistent with the standard
evolution, and a correlation of bright X-ray emission
with many of these objects suggests acceleration by 
SNR impacts (Oey 1996).  In such cases,
not all the available SN energy will be thermalized to power the shell
expansion, thus the detailed effect that discrete SNe have on the long-term
shell evolution remains unclear.  Likewise, there is evidence that the
growth rate of the shells may be overestimated, perhaps owing to an
overestimate in input wind power predicted by assumed stellar
mass-loss rates (Oey 1996; Garc\'\i a-Segura \& Mac Low 1995; Drissen
{\etal}1995).  Therefore much remains unclear about the details of 
superbubble evolution.

A demonstrative example is the superbubble DEM 152 (N44) in the Large
Magellanic Cloud.  {\it ROSAT} observations of this object clearly show
hot gas escaping through a ``blowout'' feature
(Magnier {\etal}1996), and this object also exhibits kinematics
discrepant from the standard model (Oey \& Massey 1995).  If the ISM
has low-density channels that encroach upon the superbubble 
environment, such evolutionary disruptions could signal
a significant release of hot gas into the surroundings, and not
necessarily out of the plane of galactic disks.  The details of
superbubble evolution thus have a direct consequence for the relative
volumes of coronal gas in galactic disks and halos.
Examples such as DEM 152 demonstrate the uncertainties in
our understanding of superbubble evolution, and the importance to the
global ISM.

One approach to test the assumed evolution is an
examination of the superbubble size distribution in galaxies.  This
can be predicted from the evolutionary model in combination with a
given production rate and mechanical luminosity function.  A subsequent
comparison with observed size distributions can then yield insight on
the evolutionary assumptions and possible consequences for the
inferred structure of the ISM.
The cumulative size distribution of radiative SNRs has long been used
as a diagnostic of SN parameters (\eg Hughes, Helfand, \& Kahn 1984), but
it can also be used to test evolutionary models for SNRs and
superbubbles.  An earlier investigation with a foray into
this technique was carried out for SNRs by Cioffi \& Shull (1991),
who pointed out that the differential size distribution is
a much more powerful diagnostic than the cumulative size distribution.
In this work, we present a prediction for the differential size
distribution for superbubbles and compare the results with \hi\
observations of nearby galaxies.  We will then examine the
results with regard to the structure and porosity of the ISM.


We will consider a rather rudimentary model for superbubble evolution,
with the aim of identifying dominant effects, rather than precisely
reproducing observations.  For example, the size distribution of
Sedov-Taylor SNRs is predicted to increase as $N(R) \propto R^{5/2}$
(Hughes {\etal}1984), for uniform ISM and stellar properties, whereas \hi\
hole distributions in galaxies all show size distributions that
decrease with shell radius $R$.  Our objective is to derive gross general
properties that would be expected as a result of the simplest
practical application of the standard model.  Agreements and
disagreements with observations will then be instructive in
understanding the features of the model that dominate the resultant
size distribution and shell evolution, with attendant consequences for
the structure of the ISM.

The paper is organized as follows.  In \S 2, we work out general
expressions for the assumed shell evolution and derived size distributions,
for three combinations of shell creation and mechanical power input:
(a) continuous creation, single luminosity; (b) single burst,
luminosity spectrum; and (c) continuous creation, luminosity spectrum.
In \S 3, we evaluate each of the solutions for the standard, adiabatic
shell evolution.  We also briefly examine the standard, momentum-conserving
evolution.  To obtain the slope of the mechanical luminosity spectrum,
we examine evolutionary effects in the \hii\ region luminosity
function in \S 4.  In \S 5, we compare our predictions with \hi\
observations in four nearby galaxies, with attention to the
slope, normalization, and peak of the size distributions.  We also
estimate supernova rates based on our analysis.  The consequences
are discussed in \S 6, where we examine the regimes of applicability
for our predictions.  Here we discuss the role of processes likely to
modify the size distributions, and the role of alternate shell creation
mechanisms.  These effects are likely to be dependent on galaxy
type.  We also derive expressions for the ISM porosities
resulting from our analysis, and compute porosities for the four
nearby galaxies and the Milky Way.  Our results are summarized in \S 7.

\section{General Analytic Expressions}
\subsection{Assumptions}

Superbubble expansion is presumed to be powered by the stellar winds,
and especially, SNe of the parent OB association.  It has been
customary to represent the SN power as a continuous energy injection
analogous to a wind (\eg Mac Low \& McCray 1988),
in which over half the SN energy is thermalized to drive the
shell expansion.  We will adopt this standard, adiabatic
representation (e.g., Weaver {\etal}1977; Dyson 1977).
For the purpose of comparing with \hi\ ``hole'' data and ISM
porosities, we are particularly interested in the superbubble 
cavity, which in this thin-shell approximation is considered to
evolve identically to the outer radius.  We also assume that the
mechanical luminosity $L(t)$, dominated by the SNe,
remains constant as a function of time $t$.  As shown by Shull \&
Saken (1995), this is a 
reasonable approximation for realistic slopes of the stellar initial
mass function (IMF).  This constant $L(t)$ is then assumed to
continue until the lowest-mass SN progenitors expire at $t=t_e$,
which corresponds to a stellar mass of about 8$\msol$, or a period of
roughly $t_e = 40$ Myr.  A variation in input power may occur in the
initial stages ($\la 3-4$ Myr) while $L$ is dominated by the
stellar wind phase (\eg Shull \& Saken 1995), but this stage is short
relative to $t_e$, the total period of energy input.  Leitherer \&
Heckman (1995) and Ferri\`ere (1995) show that a starburst over this
period does not substantially alter this approximation.  A
uniform ambient medium is assumed throughout, although in \S 6 we will
discuss effects expected from disk galaxy gas distributions.

As will become apparent below, the endstage of shell evolution is
vitally important to this analysis, but is fraught with uncertainties.
We will consider the following simple model, keeping in mind our 
motivation of testing the simplest-case standard evolution.  In most
cases, eventually the superbubbles' internal pressure 
$P_i\la P_0,$ the ambient pressure of the interstellar medium,
while still at ages $t< t_e$.  We will consider that for such 
objects, the superbubble growth stalls when $P_i = P_0$.
At this stage,
radiative losses are thought to become important, confining the shell
growth.  Simulations by Garc\'\i a-Segura \& Franco (1996), for
example, show that the growth of the cavity radius does stall at times
close to the sonic point.  We will assume that the superbubbles
survive at the stall 
radius $R_f$ until the input power stops at time $t_e$.  In principle, the
shells will begin to disintegrate, owing to ambient random 
motions.  However, the disintegration timescale
may be fairly long; for example, it is $R_f/v_{\rm rms} = $10 Myr for
ambient rms velocity $v_{\rm rms} = 5\ \kms$ and $R_f = 50$
pc.  Furthermore, the process of disintegration will still leave detectable
cavities, so it is unclear how this effect can be quantified.
We note that if the breakup of objects into smaller subunits occurs
such that the ratio of subunit sizes is universal for all objects,
then a power law size distribution remains unaffected (Clarke 1996).
Our analysis will therefore not consider elimination of shells at ages $t
< t_e$, nor differential elimination among the shell population.
Finally, an important feature regarding stalling is the existence of
a monotonic correspondence between $L$ and $R_f$, for a uniform ISM.

Objects that never achieve pressure
equilibrium with the ISM are simply assumed to grow until $t_e$.
Finally, all objects are assumed to survive for an additional
nominal period $t_s \ll t_e$, which is the same for all shells. 

Within individual clusters, we take the star formation to
occur in a single, instantaneous burst.  Massey \etal(1995a, b) find
that in general the duration of star formation is $\la 3$ Myr for
associations in the Magellanic Clouds and the Galaxy, which again is short
relative to $t_e$.  The spectrum of associated mechanical luminosities is
a critical parameter; we will consider both a single, global value for
$L$, and a power-law luminosity function (LF):
\begin{equation}\label{MLF}
\frac{dN}{dL} = \phi(L) = A L^{-\beta} \quad ,
\end{equation}
normalized such that $\int \phi(L)\ dL = 1$.
In \S 4 we will discuss the form of $\phi(L)$, and in particular, the
value of $\beta$; our analysis of $\phi(L)$ will assume the stellar
IMF to be constant.  We will also consider a constant cluster
formation rate $\psi(t)$, and an instantaneous burst of cluster formation. 

We first present the general expressions for the superbubble size distribution
in the case that the shells grow according to the generic law:
\begin{equation}\label{generic}
R = {\rm min}\Bigl(R(L,t),\ R_f(L)\Bigr)  \quad .
\end{equation}
where $R_f(L)$ is the radius at which a superbubble stalls, which is
dependent only on the input $L$, for uniform ISM parameters.  We now
consider the production of superbubbles for the simple cases of
luminosity distribution $\phi(L)$ and formation rate $\psi(t)$
mentioned above, with
respect to the stalling and continuous-growth evolutionary schemes.
In each case, we derive $N(R)$, defined such that the number 
of superbubbles with radii in the range $R$ to $R + dR$ is $N(R)\ dR$.
 
\subsection{Continuous Creation, Single Luminosity}

If superbubbles are generated at a constant rate $\psi$, then the number of
growing shells with radii in the range $R$ to $R + dR$ is equal to $\psi\ dt$
where $dt$ is the time interval for creation corresponding to this
radial range.  Thus, the differential size distribution is,
\begin{equation}\label{cc1l.grow}
N_{\rm grow}(R) = {\psi}\Biggl({{\partial R}\over{\partial t}}\Biggr)^{-1}
	\quad .
\end{equation}
[Note that partial derivatives with respect to $t$ and $L$ are evaluated at
constant $L$ and $t$ respectively.]

Recalling the correspondence between $L$ and stall radius $R_f$, 
the single-valued $\phi(L) = L_0$ will therefore yield a single-valued
$N_{\rm stall}\equiv N(R_f)$ for stalled objects:
\begin{equation}\label{cc1l.stall}
N_{\rm stall}(R) = \psi \Bigl(t_e-t_f(L_0)\Bigr)\cdot
	\delta\Bigl(R - R_f(L_0)\Bigr) \quad ,
\end{equation}
where $t_f(L_0)$ is the age at which a shell powered by $L_0$ stalls.

\subsection{Single Burst, Luminosity Spectrum}

If $N_b$ superbubbles are created in an instantaneous burst, then
after time $t$, the number of growing objects with radii in the range $R$ to
$R + dR$ is equal to $N_b\ \phi(L)\ dL$, where $L$ and $dL$ are the
luminosity and range corresponding to this interval in $R$.  Thus,
\begin{equation}\label{sbls.grow}
N_{\rm grow}(R) = N_b\ \phi(L)\ 
	\Biggl({{\partial R}\over{\partial L}}\Biggr)^{-1}  \quad .
\end{equation}

Similarly, the distribution of stalled shells is given by:
\begin{equation}\label{sbls.stall}
N_{\rm stall}(R) = N_b\ \phi(L)\ 
	\Biggl({{\partial R_f}\over{\partial L}}\Biggr)^{-1} \quad ,
\end{equation}

\subsection{Continuous Creation, Luminosity Spectrum}
\subsubsection{Growing Shells}

The form of $N(R)$ in this case can be obtained from the generalisation of
either equation ~\ref{cc1l.grow} or ~\ref{sbls.grow}. In the former case,
$N(R)$ is obtained by integrating equation~\ref{cc1l.grow}  over the
luminosity distribution:
\begin{equation}\label{ccls.growL}
N_{\rm grow}(R) =\int_{L_l}^{L_u} {\psi}\ \phi(L)\ 
	\Biggl({{\partial R}\over{\partial t}}\Biggr)^{-1}\ dL \quad ,
\end{equation}
whilst in the latter case $N(R)$ is obtained by integrating
equation~\ref{sbls.grow} over a continuous creation rate:
\begin{equation}\label{ccls.growt}
N_{\rm grow}(R) = \int_{t_l}^{t_u} \psi\ \phi(L)\ 
	\Biggl({{\partial R}\over{\partial L}}\Biggr)^{-1}\ dt \quad .
\end{equation}

Equations~\ref{ccls.growL} and \ref{ccls.growt} are
equivalent in this situation, provided the limits of integration are correctly
chosen.  Note that these limits are functions
of $R$, determined by the limiting conditions required to produce a
given $R$.  For a luminosity distribution truncated at 
$L_{\rm min}$ and $L_{\rm max}$, 
the lower limit of integration for equation~\ref{ccls.growL} is
\begin{equation}\label{Ll}
L_l = {\rm min}\Bigl(L(R, t_e), L_f(R)\Bigr) \quad , 
	\quad L_l \geq L_{\rm min} \quad .
\end{equation} 
Here, $L(R, t_e)$ represents the luminosity for which the shell attains
the given radius $R$ within the maximum allowable time, 
$t_e$,
whereas $L_f(R)$ is the luminosity such that a shell of luminosity
$L_f$ stalls at $R$, i.e., $R_f(L_f) = R$.
Likewise, the upper limit of integration is 
\begin{equation}\label{Lu}
L_u = L_{\rm max} \quad ,
\end{equation}
representing the luminosity of the youngest shells
that have grown to radius $R$.  In the same vein, the lower limit of
integration for equation~\ref{ccls.growt} is
\begin{equation}\label{tl}
t_l= t(R, L_{\rm max}) \quad ,
\end{equation}
implying that no shells contribute to $N(R)$ if they were formed so
recently that even the most luminous have not yet grown to $R$.  Likewise,
the upper limit is
\begin{equation}\label{tu}
t_u = {\rm min}\Bigl(t_f(L),\ t(R,L_{\rm min})\Bigr) \quad , \quad 
	t \leq t_e \quad .
\end{equation}
For the case of continuous growth with no possibility of stalling, the
most distant epoch contributing to $N(R)$ is determined only by
$L_{\rm min}$.  Thus in that case $t_u = t(R, L_{\rm min})$, as
long as this is shorter than $t_e$.  If, however, the possibility of
stalling is included, then the contribution from growing shells is
additionally limited to periods shorter than $t_f(R,L)$, where $t_f$
is the age at which a shell stalls at radius $R$ (see
equation~\ref{generic}). 

\subsubsection{Stalled Shells}

By analogy to equation~\ref{ccls.growt}, the stalled shells are
described by, 
\begin{equation}\label{ccls.stall}
N_{\rm stall}(R) = \int_{t_l}^{t_u} \psi\ \phi(L)\ 
	\Biggl(\frac{\partial R_f}{\partial L}\Biggr)^{-1} dt \quad ,
\end{equation}
with $t_l = t_f(R)$, and $t_u = t_e$, where $t_f<t_e$.

\subsubsection{Surviving Shells}

In this section, we also include the possibility of a survival phase
(\S 2.1) beyond $t_e$, of duration $t_{s}$.  For shells stalling at
$t_f < t_e$: 
\begin{equation}
N_{\rm sur}(R) = \int_{t_e}^{t_e+t_s} \psi\ dt\ \phi(L)\ 
	\Biggl(\frac{\partial R_f}{\partial L}\Biggr)^{-1} \quad ,
\end{equation}
while for those that continue growing until $t_e$, therefore having 
$t_f > t_e$:
\begin{equation}
N_{\rm sur}(R) = \int_{t_e}^{t_e+t_s} \psi\ dt\ \phi(L)\ 
	\Biggl(\frac{\partial R}{\partial L}\Biggr)^{-1} \quad .
\end{equation}
These yield,
\begin{equation}\label{ccls.sur.stall}
N_{\rm sur}(R)= \psi\ t_{s}\ \phi(L)\ 
	\Biggl({{\partial R_f}\over{\partial L}}\Biggr)^{-1} \quad 
\end{equation} 
and
\begin{equation}\label{ccls.sur.grow}
N_{\rm sur}(R)= \psi\ t_{s}\ \phi(L)\ 
	\Biggl({{\partial R}\over{\partial L}}\Biggr)^{-1} \quad .
\end{equation} 

\subsubsection{Total Size Distributions}

The complete shell evolution is summarized by 
\begin{equation}\label{bubev}
R(L,t) =\left\{ \begin{array}{ll}
           R(L,t)~~, & \quad t < t_f(L)~~,~~~t_f(L)<t_e \\
           & \\
           R_f(L)~~, & \quad t_f(L)<t<t_e+t_s~~,~~~t_f(L)<t_e \\
           & \\
           R(L,t_e)~~, & \quad t_e<t<t_e+t_s~~,~~~t_f(L)>t_e\\
           \end{array} \right\} \quad .
\end{equation}

As mentioned above, any given $L$ uniquely determines corresponding
stall parameters $R_f(L)$ and $t_f(L).$ 
Applying this relation to $t_e$, we denote the stall parameters
corresponding to $t_e$ as $R_e$ and $L_e$.  In other words, a
superbubble with input power $L_e$ would just stall at radius $R_e$
and age $t_e$.
Since the input power ceases at this time, $t_e$ is the maximum
possible stalling age.  Thus shells larger than $R_e$ are all either growing
or in the survival stage, because their $t_f > t_e$.
Conversely, the population of shells smaller than $R_e$
contains a combination of growing, stalled, and surviving objects.

Therefore, for $R < R_e$, the total distribution is given by,
\begin{equation}\label{sum<}
N(R)=N_{\rm grow}(R) + N_{\rm stall}(R) + N_{\rm sur}(R) \quad .
\end{equation}
For the case $L_{\rm min}<L(R, t_e)$ and $L_{\rm max}\rightarrow
\infty$, the contribution to $N(R)$ from growing superbubbles is
given by equation~\ref{ccls.growt}:
\begin{equation}\label{grow<}
N_{\rm grow}(R)= \int_0^{t_f(R)} \psi\ \phi(L)\ 
	\Biggl({{\partial R}\over {\partial L}}\Biggr)^{-1}\ dt \quad ,
\end{equation}
whilst the contribution from stalled systems  is given by
equation~\ref{ccls.stall}:
\begin{equation}\label{stall<}
N_{\rm stall}(R) = \psi \cdot \Bigl(t_e - t_f(R)\Bigr)\ \phi(L)\ 
	\Biggl({{\partial R_f}\over{\partial L}}\Biggr)^{-1} \quad .
\end{equation}
The distribution for surviving shells, $N_{\rm sur}(R)$, is given by
equation~\ref{ccls.sur.stall}.

For $R > R_e$, the total size distribution is,
\begin{equation}\label{sum>}
N(R)=N_{\rm grow}(R) + N_{\rm sur}(R) \quad .
\end{equation}
For the contribution due to growing superbubbles,
equation~\ref{ccls.growt} is now written:
\begin{equation}\label{grow>}
N_{\rm grow}(R) = \int_0^{t_e} \psi\ \phi(L)\ 
	\Biggl({{\partial R}\over{\partial L}}\Biggr)^{-1}\ dt \quad ,
\end{equation}
again for $L_{\rm min}<L(R, t_e)$ and $L_{\rm max}\rightarrow\infty$.
Likewise, the shells in the survival phase are, by assumption,
preserved at their instantaneous radii at time $t_e$, so that $N_{\rm
sur}(R)$ is given by equation~\ref{ccls.sur.grow}.

\section{Expressions Specific to the Standard Evolution}

We now incorporate the standard evolution for adiabatic, pressure-driven
superbubbles to the expressions developed above.  The growth of the
shell radius is described by (\eg Weaver {\etal}1977):
\begin{equation} \label{Rcgs}
R = \Biggl({250\over {308\pi}}\Biggr)^{1/5} L^{1/5} \rho^{-1/5}\
	t^{3/5}  \quad ,
\end{equation}
where $\rho$ is the mass density of a uniform ambient
medium, and all units are cgs.
The corresponding evolution of the interior pressure is
\begin{equation}\label{Picgs}
P_i = {7\over{(3850\pi)^{2/5}}}\ L^{2/5}\ \rho^{3/5}\ t^{-4/5} \quad .
\end{equation}
In what follows, it is convenient to use $R_e,\ t_e,$ and $L_e$ as
scaling parameters, recalling that $R_e$ and $L_e$ correspond to the
stall criterion at the characteristic time $t_e$.
Equations~\ref{Rcgs} and \ref{Picgs} can now be written, 
\begin{equation} \label{R2}
\frac{R}{R_e} = 
	\Biggl(\frac{L}{L_e}\Biggr)^{1/5}
	\Biggl(\frac{t}{t_e}\Biggr)^{3/5}  \quad ,
\end{equation}
\begin{equation}\label{P2}
\frac{P_i}{P_0} = 
	\Biggl(\frac{L}{L_e}\Biggr)^{2/5}
	\Biggl(\frac{t}{t_e}\Biggr)^{-4/5} \quad .
\end{equation}

As discussed previously, we consider that the shell growth stalls when
$P_i = P_0,$ the ambient pressure.  This condition therefore implies a
stall age: 
\begin{equation}\label{tfL}
\frac{t_f}{t_e} = 
	\Biggl(\frac{L}{L_e}\Biggr)^{1/2} \quad ,
\end{equation}
Thus, by equation~\ref{R2},
\begin{equation}\label{Rf} 
\frac{R_f}{R_e} = 
	\Biggl(\frac{L}{L_e}\Biggr)^{1/2} \quad .
\end{equation}
Equations \ref{tfL} and \ref{Rf} demonstrate the correspondence between
stalling age $t_f$ and radius $R_f$ for a given input power $L$.
The corresponding relation between $t_f$ and $R_f$ is,
\begin{equation}\label{tfRf}
\frac{t_f}{t_e} = 
	\frac{R_f}{R_e} \quad .
\end{equation}
Thus the final age $t_f$ is directly proportional to the final radius $R_f$.  

For a given $t_e$, equations~\ref{Rcgs} and \ref{Picgs} give:
\begin{equation}\label{Re}
R_e = \frac{5}{7^{1/2}}\ \Bigl(\mu m_{\rm H}\Bigr)^{-1/2} n^{-1/2}\ 
	P_0^{1/2}\ t_e
\end{equation}
\begin{equation}\label{Le}
L_e = \frac{550\pi}{7^{3/2}}\ (\mu m_{\rm H})^{-3/2}n^{-3/2}\ 
	P_0^{5/2}\ t_e^2 \quad ,
\end{equation}
in cgs units, where $n$ is the number density of the uniform ambient
medium and $\mu$ is the mean particle weight.  Note that
equations~\ref{Re} and \ref{Le} are applicable in general for the
relation between $R_f$, $t_f$, and $L_f$ as well.  For ISM parameters 
$n = 0.5\ \cc,\ \mu=1.25$, and $P_0=3\times 10^{-12}\ {\rm dyne\
cm^{-2}}$, the characteristic time $t_e = 40$ Myr implies
$R_e = 1300$ pc and $L_e = 2.2\times 10^{39}\ergs$.

\subsection{Continuous Creation, Single Luminosity}
 
As described in \S2.2, the growing shells are the interesting case
here, for which equations~\ref{cc1l.grow} and \ref{R2} imply,
\begin{equation}\label{E.cc1l.grow}
N_{\rm grow}(R) = \frac{5}{3}\ \psi\ \frac{t_e}{R_e}
	\Biggl(\frac{L_0}{L_e}\Biggr)^{-1/3}
	\Biggl(\frac{R}{R_e}\Biggr)^{2/3} \quad .
\end{equation}
Thus for a single-valued LF of luminosity $L_0$, the size distribution
is an increasing 
function of $R^{2/3}$, which results from the slower growth of large
shells.  As the stalled shells accumulate, they also superimpose a
$\delta$-function at $R_f(L_0)$, described by equation~\ref{cc1l.stall}.

\subsection{Single Burst, Luminosity Spectrum}

Here we assume that $\phi(L)$ is a power law as given by equation~\ref{MLF}.
For the growing shells with age $t<t_e$, equations~\ref{sbls.grow} and
\ref{R2} yield, 
\begin{equation}\label{E.sbls.grow}
N_{\rm grow}(R) = 5AN_b(1-F_{\rm st})\ \frac{L_e^{1-\beta}}{R_e}
	\Biggl(\frac{R}{R_e}\Biggr)^{4-5\beta}
	\Biggl(\frac{t}{t_e}\Biggr)^{-3+3\beta} \quad ,
\end{equation}
where $1-F_{\rm st}$ is the fraction of $N_b$ corresponding to growing
shells.  The fraction of stalled shells is given by, 
\begin{equation}\label{Fcoeff}
F_{\rm st} = \int_{L_{\rm min}}^{L} A L^{-\beta}\ dL \quad ,
\end{equation}
where $L$ is the luminosity corresponding to the largest stalled shells.
As we shall see in \S 5, realistic values of the LF
index fall in the range $1\la\beta\la 3$, for which
$N(R)\propto R^{4-5\beta}$ would be an inverse power
law in $R$.  This results from the larger fraction of small, weak-$L$
shells. 

The stalled shells are described by equation~\ref{sbls.stall}.
Together with equation~\ref{Rf}, this gives,
\begin{equation}\label{E.sbls.stall}
N_{\rm stall}(R) = 2AN_b F_{\rm st}\ \frac{L_e^{1-\beta}}{R_e}
	\Biggl(\frac{R}{R_e}\Biggr)^{1-2\beta} \quad .
\end{equation}

The total size distribution at times before all the objects have
stalled and $R<R_e$ would be the sum of equations~\ref{E.sbls.grow} and
\ref{E.sbls.stall}.  Owing to the large numbers of weak-$L$ objects,
equation~\ref{E.sbls.stall} will normally dominate $N(R)$ in relative
numbers.  However, these stalled shells will only be present out to
radius $R_f(t_b)$
corresponding to the age $t_b$ of the burst.  At this radius, there will be
a discontinuous jump by a factor of $\frac{5}{2}$, resulting from the
different coefficients of $\frac{\partial R}{\partial L}$
(cf. equation~\ref{R2}) and $\frac{\partial R_f}{\partial L}$
(cf. equation~\ref{Rf}).  The subset of objects with $R>R_e$ will not
stall in $t<t_e$ and will therefore follow the relation given by
equation~\ref{E.sbls.grow}.

\subsection{Continuous Creation, Luminosity Spectrum}
\subsubsection{$R<R_e$}

We shall now evaluate the expressions in \S 2.4.4, for the population
of shells with radii $R<R_e$, incorporating $\phi(L)$ as given by
equation~\ref{MLF}.  For $\beta>\frac{2}{3},\ L_{\rm min}<L(R, t_e),$
and $L_{\rm max}\rightarrow\infty,$ the size distribution for the
growing shells is given by equation~\ref{grow<}. 
With the aid of equation~\ref{R2} this yields,
\begin{equation}\label{E.grow<.>b}
N_{\rm grow}(R) = 5A\psi\ \frac{L_e^{1-\beta}}{R_e} \ \frac{t_e}{-2+3\beta}\ 
	\biggl(\frac{R}{R_e}\biggr)^{4-5\beta}\ 
	\biggl(\frac{t_f}{t_e}\biggr)^{-2+3\beta} \quad .
\end{equation}
For each $R$ we are summing over all ages up to $t_f$ that yield
a stall radius $R$, so that (from equation~\ref{tfRf})
equation~\ref{E.grow<.>b} becomes,
\begin{equation}\label{E.grow<.b>}
N_{\rm grow}(R) = 5A\psi\ \frac{L_e^{1-\beta}}{R_e} \ 
	\frac{t_e}{-2+3\beta}\ \biggl(\frac{R}{R_e}\biggr)^{2-2\beta} \quad .
\end{equation}


In the case that $\beta < \frac{2}{3}$, equation~\ref{ccls.growt} is
dominated by $t_l$ (equation~\ref{tl}), and if $L_{\rm min} > L(R,
t_e)$, then $t_u = t(R, L_{\rm min})$.  For both of these cases, 
\begin{equation}\label{E.grow<.b<}
N_{\rm grow}(R) = 5A\psi\  \frac{L_e^{1-\beta}}{R_e}\
	\biggl(\frac{L_{\rm max}}{L_e}\biggr)^{\frac{2}{3}-\beta}\ 
	\frac{-t_e}{-2+3\beta}\  
	\biggl(\frac{R}{R_e}\biggr)^{2/3} \quad .
\end{equation}
Thus for $L_{\rm min}\neq 0$, the slope of the
size distribution turns over from $4-5\beta$ in equation~\ref{E.grow<.>b} to a
positive slope of $\frac{2}{3}$, at $R<R(L_{\rm min},t_e)$, where
$R(L_{\rm min}, t_e)$ corresponds to the case in the absence of
stalling.  Beware that the 
resulting peak in $N_{\rm grow}(R)$ does not correspond to the maximum
peak in the total $N(R)$, since the latter is dominated by stalled
shells, as we shall see below.

For the stalled shells, equations~\ref{stall<} and \ref{Rf} yield,
\begin{equation}\label{E.stall<}
N_{\rm stall}(R) = 2A\psi\ \frac{L_e^{1-\beta}}{R_e}\ t_e\ 
	\biggl(\frac{R}{R_e}\biggr)^{1-2\beta}
	\biggl(1-\frac{R}{R_e}\biggr) \quad .
\end{equation}
Note that for these stalled objects, the term
$\Bigl(1-\frac{R}{R_e}\Bigr)= \Bigl(1-\frac{t_f}{t_e}\Bigr)$
by equation~\ref{tfRf}.  This therefore represents the fraction of
their lifetime $t_e$ that is spent in the stalled state, for shells of
radius $R$.  Although multiplying out equation~\ref{E.stall<} yields
two terms with dependences $R^{1-2\beta}$ and $R^{2-2\beta},\ N_{\rm
stall}(R)$ is dominated by $R^{1-2\beta}$ since we are in the
regime $\frac{R}{R_e}<1$.  This is the same dependence
as the stalled objects in equation~\ref{E.sbls.stall}.

The shells in the survival phase are given by
equation~\ref{ccls.sur.stall}, which becomes
\begin{equation}\label{E.ccls.sur.stall}
N_{\rm sur}(R) = 2A\psi\ \frac{L_e^{1-\beta}}{R_e} \ t_s\ 
	\biggl(\frac{R}{R_e}\biggr)^{1-2\beta} \quad .
\end{equation}
Again this is similar to equation~\ref{E.sbls.stall}, since this case
is equivalent to the distribution resulting from a single burst
containing $\psi t_s$ objects.

Adding together equations~\ref{E.grow<.b>}, \ref{E.stall<}, and
\ref{E.ccls.sur.stall}, the total size distribution for superbubbles with
$R < R_e$ and $\beta > \frac{2}{3}$ is:
\begin{equation}\label{total<}
N(R) = A\psi\ {\frac{L_e^{1-\beta}}{R_e~}}~
	\biggl({\frac{R}{R_e}}\biggr)^{1-2\beta}\
	\Biggl[2\bigl(t_e+t_s\bigr) +
	{\frac{~9-6\beta}{-2+3\beta}}\ t_e\ 
	\biggl({\frac{R}{R_e}}\biggr) \Biggr] \quad .
\end{equation}

\subsubsection{$R>R_e$}

For shells larger than the characteristic radius $R_e$, the
distribution of growing objects is given by equation~\ref{grow>}.
This is identical to equation~\ref{grow<}, except that the 
integration is over the range (0, $t_e$) rather than (0, $t_f(L)$).
For $\beta>\frac{2}{3}$, equation~\ref{grow>} therefore yields,
\begin{equation}\label{E.grow>}
N_{\rm grow}(R) = 5A\psi\ \frac{L_e^{1-\beta}}{R_e}\ 
	\frac{t_e}{-2+3\beta}\ \biggl(\frac{R}{R_e}\biggr)^{4-5\beta}\quad .
\end{equation}
Note that the $R$ dependence of $N_{\rm grow}(R)$ in
equation~\ref{E.grow>} is different from that of shells that grow
with the possibility of stalling (equation~\ref{E.grow<.b>}).  This is
due to the fact that the upper limit of integration in
equation~\ref{ccls.growt} is a function of
$R$ only when the possibility of stalling is included.

The shells in the survival stage are described by
equation~\ref{ccls.sur.grow}, yielding,
\begin{equation}\label{E.ccls.sur.grow}
N_{\rm sur}(R) = 5A\psi\ \frac{L_e^{1-\beta}}{R_e}\ t_s\ 
	\biggl(\frac{R}{R_e}\biggr)^{4-5\beta} \quad .
\end{equation}

Adding together equations~\ref{E.grow>} and \ref{E.ccls.sur.grow}, the
overall size distribution for supergiant shells with $R> R_e$ and
$\beta>\frac{2}{3}$ is:
\begin{equation}\label{total>}
N(R) = 5A\psi\ \frac{L_e^{1-\beta}}{R_e} 
	\biggl(\frac{R}{R_e}\biggr)^{4-5\beta}\ 
	\biggl[\frac{t_e}{-2+3\beta} + t_s\biggr] \quad .
\end{equation}
This again exhibits the $R^{4-5\beta}$ dependence as seen in
equation~\ref{E.sbls.grow}, which describes the distribution for freely
growing shells.  This reflects the fact that objects in this size range 
will never attain the stalling criterion.
In general, this relation is a very steep function of $R$ compared to
equation~\ref{total<} for $R<R_e$.  The criterion $R>R_e$
corresponds to $L>L_e$, which may provide a convenient defining
criterion for these larger-scale phenomena, for example, starburst
events.  There might possibly be applications for
equation~\ref{total>} for a large sample of starburst phenomena with
uniform properties.

In the case that $\beta<\frac{2}{3}$, equation~\ref{ccls.growt} is
dominated by the lower limit $t_l$, as in the regime for $R<R_e$.  
$N_{\rm grow}(R)$ in that case is therefore given by equation~\ref{E.grow<.b<}.
Note that, for $t_{s} > 0,\ N(R)$ is {\it not} continuous
at $R=R_e$.  This is a consequence of the change 
in slope of the $L-R$ relation for shells surviving at $R_e$, as is
apparent from equations~\ref{R2} and \ref{Rf} (see \S 2.4.3).

\subsection{Robustness of the Result}

The dependence $N(R)\propto R^{1-2\beta}$ appears to be a fairly
robust description, since stalled superbubbles quickly
dominate most situations for both constant and instantaneous
superbubble formation.  In addition, we also mentioned above that the slope
will not change if the breakup of shells into subunits is independent
of the original size (Clarke 1996).

Furthermore, the $R^{1-2\beta}$ dependence turns out not to be unique
to the standard evolution.  Consider the general relation,
\begin{equation}\label{Rgen}
\Biggl(\frac{R}{R_e}\Biggr) = \Biggl(\frac{L}{L_e}\Biggr)^x\ 
	\Biggl(\frac{t}{t_e}\Biggr)^y \quad .
\end{equation}
Within factors of unity, the stalling criterion is equivalent to setting
the shell ram pressure $\rho_s v^2 \sim P_0$, where $\rho_s$ is the
shell density and $v$ is the shell expansion velocity.  Therefore, this is
the same as requiring that $v$ stall at a limiting velocity $v_0$
associated with $P_0$ of the ambient medium.  We have, 
\begin{equation}
v = \frac{dR}{dt} = y\ \frac{R_e}{t_e}\ \biggl(\frac{L}{L_e}\biggr)^x\
	\biggl(\frac{t}{t_e}\biggr)^{y-1} \quad .
\end{equation}
Setting this equal to $v_0$, we obtain,
\begin{equation}
\frac{t_f}{t_e} = \biggl(\frac{v_0}{y}\cdot\frac{t_e}{R_e}\biggr)^
	\frac{1}{y-1}\ \biggl(\frac{L}{L_e}\biggr)^\frac{x}{1-y} \quad ,
\end{equation}
which is the generalization of equation~\ref{tfL}.  Note that $R_e$ is
defined such that $\frac{v_0}{y}\cdot\frac{t_e}{R_e}=1$.  With
equation~\ref{Rgen}, we then have,
\begin{equation}
\frac{t_f}{t_e}=\frac{y}{v_0}\cdot\frac{R_e}{t_e}\ 
	\biggl(\frac{R_f}{R_e}\biggr) \quad ,
\end{equation}
analagous to equation~\ref{tfRf}.  The relation
$t_f\propto R_f$ therefore applies generally to any growth law
described by equation~\ref{Rgen}.  In the standard adiabatic model,
$x=\frac{1}{5}$ and $y=\frac{3}{5}$, yielding $t_f\propto L^{1/2}$.  
Other values of $x$ and $y$ yielding $\frac{x}{1-y}=\frac{1}{2}$ will
also produce equations~\ref{tfL} -- \ref{tfRf}, and therefore
our previous expressions for $N_{\rm stall}(R)$ (equation~\ref{E.stall<})
and $N_{\rm sur}(R)$ (equation~\ref{E.ccls.sur.stall}) would still be
valid.

\subsubsection{Momentum-Conserving Evolution}

It turns out that the non-adiabatic, momentum-conserving shell
evolution described by Steigman, Strittmatter, \& Williams (1975)
fulfills this condition.  The momentum-conserving growth is given by,
\begin{equation}\label{Rmc}
\Biggl(\frac{R}{R_e}\Biggr) = \Biggl(\frac{L}{L_e}\Biggr)^{1/4}
	\Biggl(\frac{t}{t_e}\Biggr)^{1/2} \quad ,
\end{equation}
Therefore, the only component of $N(R)$ that differs from the
adiabatic case is that for
$N_{\rm grow}(R)$, for which equations~\ref{grow<} and \ref{Rmc} give:
\begin{equation}\label{E.grow<.mc}
N_{\rm grow}(R) = 4A\psi\ \frac{L_e^{1-\beta}\ t_e}{R_e}\ 
	\frac{1}{-1+2\beta}\ \Biggl(\frac{R}{R_e}\Biggr)^{2-2\beta} \quad ,
\end{equation}
in contrast to equation~\ref{E.grow<.b>}.  The total size distribution
for the momentum-conserving evolution in the regime 
$R<R_e,\ \beta>\frac{1}{2}$ is therefore,
\begin{equation}\label{total<.mc}
N(R) = A\psi\ {\frac{L_e^{1-\beta}}{R_e~}}~
	\biggl({\frac{R}{R_e}}\biggr)^{1-2\beta}
	\Biggl[2\bigl(t_e+t_s\bigr) +
	{\frac{~6-4\beta}{-1+2\beta}}\ t_e\ 
	\biggl({\frac{R}{R_e}}\biggr) \Biggr]
	\quad ,
\end{equation}
which may be compared to equation~\ref{total<}.  $N(R)$ in this case
(equation~\ref{total<.mc}) is still dominated by the term having
$R^{1-2\beta}$ dependence.

To check whether typical observed superbubbles are still in the regime
$R<R_e$ for this case, we need to compute the new value of $R_e$, again
for $t_e = 40$ Myr.  The unscaled relations for the radius and
interior pressure are:
\begin{equation}
R = \biggl(\frac{3L}{\pi\rho v_\infty}\biggr)^{1/4}\ t^{1/2}
\end{equation}
\begin{equation}
P_i = \frac{\Dt{M}v_\infty}{4\pi R^2} \quad ,
\end{equation}
where $\Dt{M}$ and $v_\infty$ are the mass-loss rate and velocity of
the ``wind,'' thereby giving $L = \frac{1}{2}\Dt{M} v_\infty^2$.  
The pressure equilibrium stalling criterion now yields,
\begin{equation}\label{Remc}
R_e = 6^{1/2}\ \Bigl(\mu m_{\rm H}\Bigr)^{-1/2}\ n^{-1/2}\ 
	P_0^{1/2}\ t_e \quad .
\end{equation}
The dependences are the same as equation~\ref{Re}, but with a
coefficent of $6^{1/2}$ in the momentum-conserving case versus
$5/7^{1/2}$ in the adiabatic case.  Thus, for a given stall age $t_f$,
the corresponding stall radius of a momentum-conserving superbubble
will be larger than that of an adiabatic superbubble.  This results
from the fact that the same value of $t_f$ corresponds to a larger
value of $L$ in the momentum-conserving case because more energy
is needed to maintain growth during the entire period $t_f$.
Therefore the final stall radius $R_f$ for the momentum-conserving
evolution is larger than the adiabatic case for a given $t_f$.  For the same
default ISM parameters as the adiabatic case, equation~\ref{Remc} yields $R_e =
1700$ pc.  Most observed superbubbles in galaxies indeed have radii
smaller than this value.  The form $N(R)\propto R^{1-2\beta}$ is
therefore a fairly robust result for this simple representation 
of superbubble evolution.  

\section{Deriving the Mechanical LF from the \hii\ Region LF}

We must now consider the form of $\phi(L)$, the mechanical LF
(MLF; equation~\ref{MLF}), with regard to its slope $\beta$ and the upper and
lower limits.  The MLF results from the stellar census
present in the OB associations, which in turn manifests its presence
through the \hii\ region LF (\hii\ LF).  As recognized by Heiles
(1990), the MLF and \hii\ LF are closely linked.  We will focus on clusters
that are sufficiently rich so that  statistical fluctuations in stellar
membership have a negligible effect on the mean \Ha\ luminosity per star.
We term such clusters ``saturated,'' and consider
this limit for now, deferring 
a discussion of smaller, ``unsaturated'' clusters until \S 4.3.

For the saturated clusters,
the total initial \Ha\  luminosity ${\Lha}_0$, and the mechanical luminosity
$L$, are both proportional to the total number of stars in the
cluster, thus implying $L\propto {\Lha}_0$. 
Therefore, the slope of the MLF is equal to that
of the \hii\ LF.  This is not, however,
necessarily the same as the slope of the 
{\it observed} \hii\ LF (von Hippel and Bothun 1990), because the
\hii\ region luminosity fades with time as the ionizing stars expire,
and the observed \hii\ LF contains objects of differing ages.  Therefore,
before we simply take the empirical slope of the \hii\ LF as that of the
MLF, we must first consider the possible effect of evolution
on the \hii\ LF.  We will now examine this effect analytically.  

\subsection{Analytic Evolution of the \hii\ Region LF}

We characterise the evolution of the \hii\ region luminosity by 
\begin{equation}\label{f}
{\Lha}={\Lha}_0 f(t)
\end{equation}
where $f(0) = 1$.  We thus assume that the fractional fading in a
given time, and hence the IMF, is the same for all clusters.
Since the observed \hii\ LFs are well described as power laws (\eg
Kennicutt, Edgar, \& Hodge 1989), we will presume the initial \hii\
LF to have the same form: 
\begin{equation}\label{HLF}
\frac{dN_{\ha}}{d{\Lha}_0} \equiv \Phi({\Lha}_0) 
	= A_0 {\Lha}_0^{-\beta} \quad .
\end{equation}
We adopt the index $\beta$ as in the previous sections, because as
argued above, we assume that $\phi(L)$ has the same exponent as
$\Phi({\Lha}_0)$. 
As with $\phi(L)$ (equation~\ref{MLF}), $\Phi({\Lha})$ is defined
such that $\Phi({\Lha})\ d{\Lha}$ is the fraction of objects with \Ha\
luminosities in the range ${\Lha}$ to ${\Lha} + d{\Lha}$.
If the magnitude of $\frac{df}{dt}$ is large at early times, we would
expect the steady-state observed \hii\ LF to be steeper than the
initial, since the brightest objects are quickly diminished.  On the
other hand, if $f$ evolves strongly at late times, then the initial
slope $\beta$ is largely preserved.

The evolution of $\Phi({\Lha})$ may be considered analogous to that
of $N(R)$ (equation~\ref{ccls.growL}) above:
\begin{equation}
 \Phi({\Lha})= \int_{\Lha}^{{\Lha}_{\rm 0,up}} \psi\ \Phi({\Lha}_0)\ 
	\biggl(-{{\partial{\Lha}}\over{\partial t}}\biggr)^{-1} 
	d{\Lha}_0 \quad ,
\end{equation}
where $\psi$ is the constant cluster formation rate as before and the partial
derivative with respect to $t$ is evaluated at constant ${\Lha}$.
The observed \hii\ LF is determined by the contribution of objects at
all ages to a given luminosity bin, and these therefore
fall in the range $({\Lha}, {\Lha}_{\rm 0, up})$.  From equation~\ref{f}, 
\begin{equation}
\frac{\partial {\Lha}}{\partial t} = {\Lha}_0\ \frac{df}{dt} \quad .
\end{equation}
So in general,
\begin{equation}
\Phi({\Lha}) =  \int_{{\Lha}}^{{\Lha}_{\rm 0,up}}  \psi\ 
	\Phi({\Lha}/f)\ 
	\biggl(-{\Lha}_0\ \frac{df}{dt}\biggr)^{-1}\ d{\Lha}_0 \quad 
\end{equation}
where $f$ and $\frac{df}{dt}$ are functions of ${\Lha}$ and ${\Lha}_0$
as described by equation~\ref{f}.
The upper limit of integration
is determined by the upper limit to the initial \hii\ LF, 
${\Lha}_{0,\rm up}$.  For a power law LF given by equation~\ref{HLF}:
\begin{equation}\label{evol.gen}
\Phi({\Lha}) = -A_0 {\Lha}^{-\beta}\ \int_{{\Lha}/{\Lha}_{0,\rm up}}^1 
	\psi(t)\ f^{\beta-1} \biggl(\frac{df}{dt}\biggr)^{-1}\ df \quad .
\end{equation}

We now consider the form of $f(t)$.  This depends on the stellar mass
($m$) vs. main-sequence lifetime ($t_{\rm ms}$) relation, as well as the
relative contributions to total ${\Lha}$ from stars of  different 
masses.   Leitherer (1990, Figure~8) shows that fortuitously, for a
standard Salpeter (1955) IMF, the contribution to ${\Lha}$ 
from each unit mass bin $\frac{d{\Lha}}{dm}$, is approximately
constant for all $m > 20\msol$, while for lower-mass stars it 
plummets steeply.  

As an exploratory case, we  therefore
make the approximation  that only stars with $m \geq 20\msol$ contribute to
${\Lha}$, and that the different masses contribute equally.
We represent the $m - t_{\rm
ms}$ relation for $m \geq 20\msol$ 
as a power law:
\begin{equation}\label{mtms}
t_{\rm ms} \propto m^{-d} \quad .
\end{equation}
The Geneva stellar evolutionary models (Schaerer {\etal}1993; Maeder
1990; Maeder \& Meynet 1988) show that this relation actually
increases somewhat 
faster than a simple power law, but for our purposes, the
approximation is adequate.  Fits to their models, which are the same
as those used in Leitherer's (1990) study, yield $d\sim 0.65$ in
the range $20 \leq m \leq 120\msol$.
If we furthermore assume that stars in each mass bin contribute a constant
luminosity for the duration of their main sequence lifetimes and nothing
thereafter, we then obtain the fading function,
\begin{equation}\label{cathie.f} 
f = \left\{ \begin{array}{ll}
	1 ~~, & \quad t < t_i \\
	& \\
	\frac{(t/t_{20})^{-1/d} - 1}{(t_i/t_{20})^{-1/d} - 1} ~~, 
	& \quad t \geq t_i
	\end{array} \right.
\end{equation}
where $t_{20} \equiv t_{\rm ms}(20\msol)$, and $t_i$ is the time at
which the ${\Lha}$ fading begins, which is $t_{\rm ms}(m_{\rm up})$,
where $m_{\rm up}$ is the upper-mass limit of the IMF.
Applying equation~\ref{cathie.f} to equation~\ref{evol.gen}, we obtain:
\begin{equation}\label{cathie.phi}
\Phi({\Lha}) = A_0\psi\ {\Lha}^{-a}\ 
	\Biggl[\int_{{\Lha}/{\Lha}_{0,\rm up}}^1\ 
	\frac{\Bigl(\frac{m_{\rm up}}{m_{20}}-1\Bigr)f^{\beta-1}t_{20} d}
	{\Bigl[\Bigl(\frac{m_{\rm up}}{m_{20}}-1\Bigr)f 
	+1\Bigr]^{d+1}}\ df\ + t_i \Biggr] \quad ,
\end{equation}
using $a$ to denote the slope of the observed \hii\ LF, 
$\Phi({\Lha})\propto {\Lha}^{-a}$.
The last term in equation~\ref{cathie.phi} corresponds to shells with
ages $t<t_i$, for which equation~\ref{evol.gen} diverges.

\begin{figure*}
\begin{center}
   \epsfbox{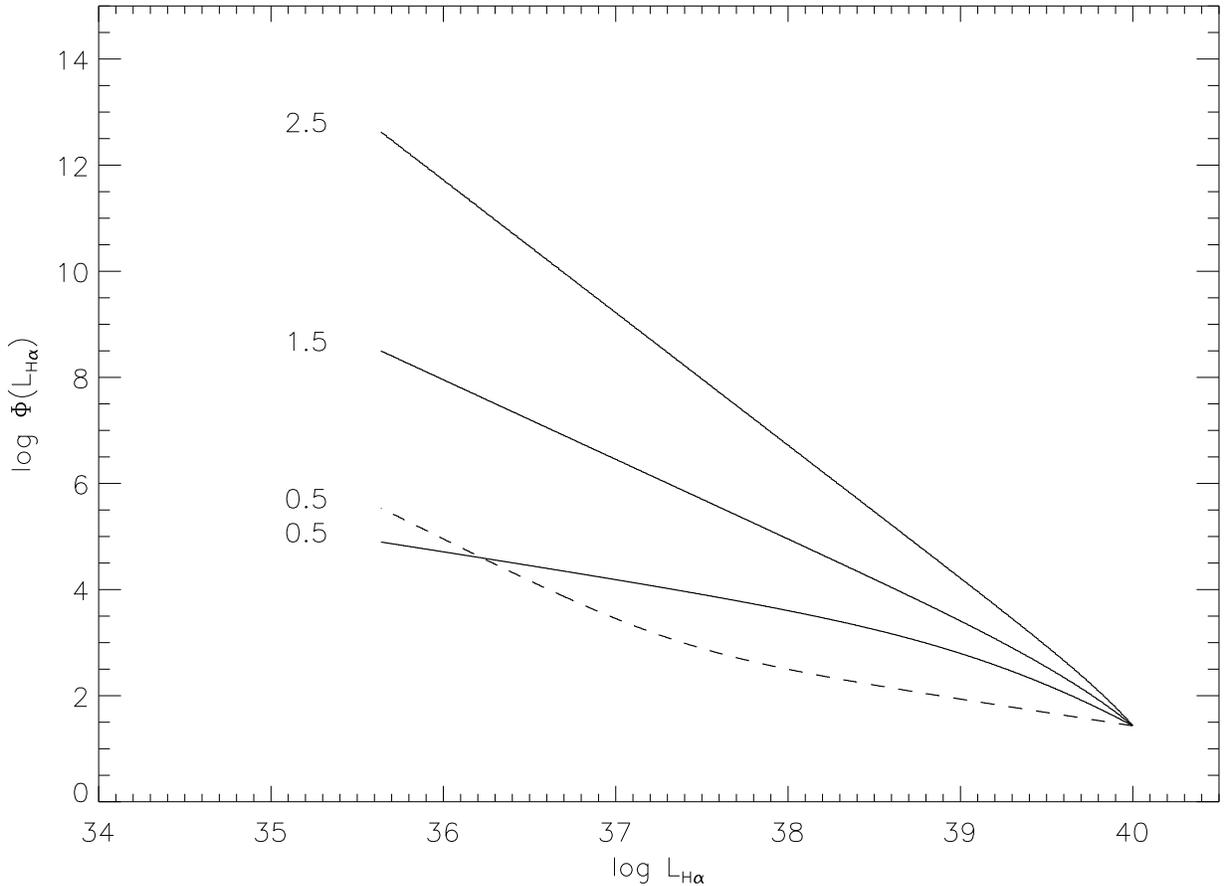}
\end{center}
\caption[]{\hii\ region LF resulting from the fading function given by
equation~\ref{cathie.f}, for $\log ({\Lha}_{\rm up}/\ergs) = 40,\ 
d=0.65$, and $m_{\rm up}=120\ \msol$  
(solid lines) and $m_{\rm up}=10^7\ \msol$ (dashed line).  The input
values of $\beta$ are shown for each model.
\label{F.cathie.phi}}
\end{figure*}

Equation~\ref{cathie.phi} demonstrates that for $\beta > d + 1$, the
integral is always dominated by its upper limit, so that ageing does not
affect the form of the \hii\ LF in this case. For $\beta < d + 1$,  if
$\frac{m_{\rm up}}{m_{20}}$ is sufficiently large, then $\Phi({\Lha})$
will asymptotically approach a slope of $d+1$. 
Values for the observed slope are found to be $a\sim
2\pm 0.5$ (\eg Kennicutt {\etal}1989).  Figure~\ref{F.cathie.phi}
shows the resultant $\Phi({\Lha})$ from
equation~\ref{cathie.phi}, numerically integrated for initial 
slopes $\beta$ ranging from 0.5 to 2.5, with $d = 0.65$, ${\Lha}_{0,\rm
up} = 10^{40}\ \ergs$, and $m_{\rm up} = 120\msol$.  The curves are
normalized to the same upper value to facilitate comparison.  
Each curve is labeled with the input value of $\beta$, which is within
4\% of the resultant slope 
$a$, fitted over  the region $\log{\Lha}<36.5$.  Thus it is apparent that
evolution has essentially no effect on these slopes for $m_{\rm up} 
=120\msol$.  With the dashed line, we also show the result for $\beta
= 1$, $d=0.65$ and $m_{\rm up} = 10^7\ \msol$; this model confirms that 
the observed slope does tend to the asymptotic limit
of $d+1=1.65$ if $m_{\rm up}$ is sufficiently large, although this
effect is clearly negligible for realistic situations.

Alternatively, we can estimate $f(t)$ directly from the population
synthesis work of Leitherer \&
Heckman (1995).  Figure~37 of that work shows the evolution of the
total ionizing photon emission rate $S(\rm H^0)$, and 
hence ${\Lha}$,  for clusters with different IMFs and $m_{\rm up}$. The
fading function is relatively insensitive to
these variables and is well approximated by a power law decrease with time,
which does not drop to zero at $t_{20}$, but persists with the same
power law at late times.  From the Leitherer \& Heckman (1995)
standard model with a Salpeter IMF of slope 2.35 and 
$m_{\rm up} = 100\msol$, we fit a fading function slope $\eta = 5.0$
in the regime $\log (t/{\rm Myr}) > 6.5$ and $\log \bigl(S(\rm H^0)/{\rm
s^{-1}}\bigr) > 44$.  This slope was also found by Beltrametti
{\etal}(1982). 

\begin{figure*}
\begin{center}
   \epsfbox{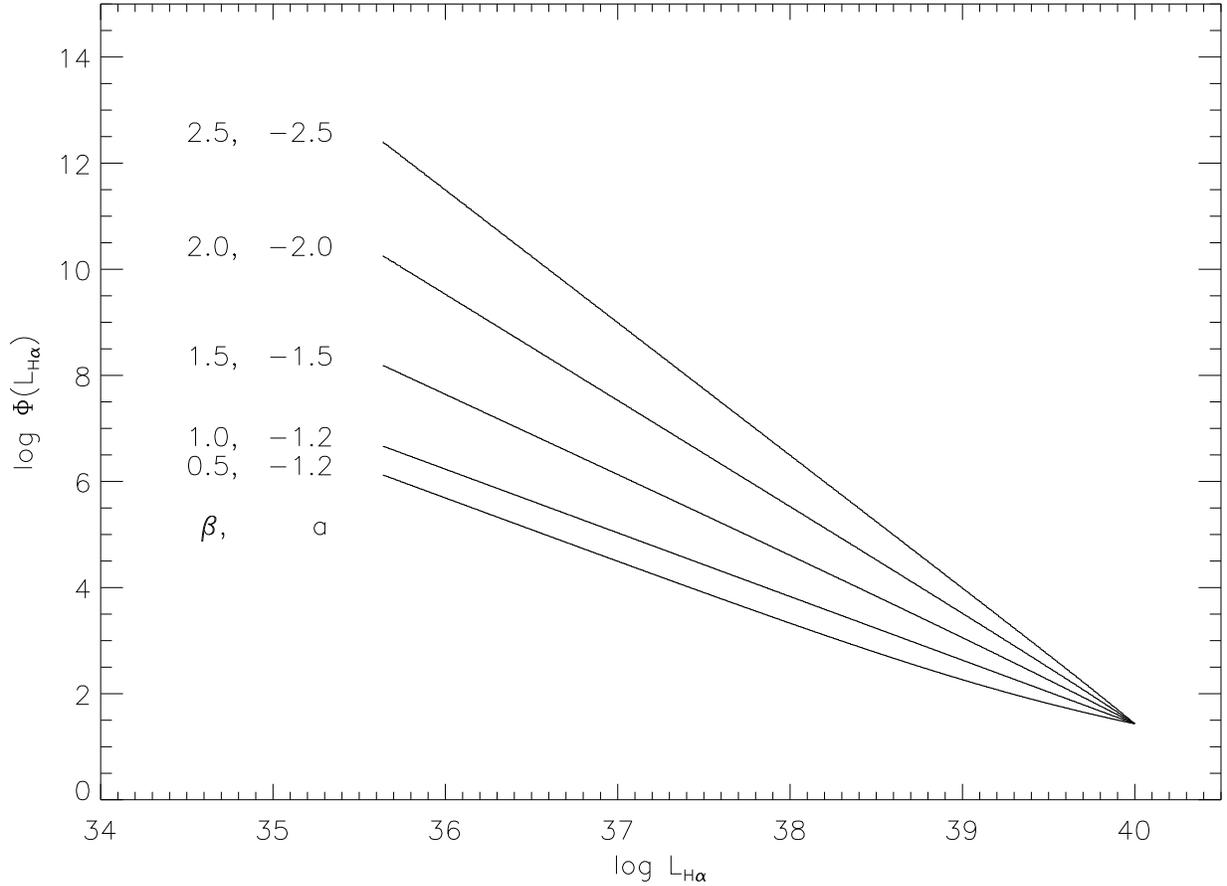}
\end{center}
\caption[]{\hii\ region LF resulting from the fading function given by
equation~\ref{lh95.f}, for $\log ({\Lha}/\ergs) = 40,\ \eta = 5.0,$
and $m_{\rm up}=120\ \msol$.  The input 
values of $\beta$ are shown for each model, along with the resulting
slope $a$, fitted to the region $\log {\Lha}<36.5$.
\label{F.lh95.phi}}
\end{figure*}

We thus parameterize the evolution as:
\begin{equation}\label{lh95.f}
f = \left\{ \begin{array}{ll}
	1 ~~, & \quad t < t_i \\
	& \\
	(t/t_i)^{-\eta} ~~, & \quad t \geq t_i
	\end{array} \right.
\end{equation}
so that integration of
equation~\ref{evol.gen} yields:
\begin{equation}\label{lh95.phi}
\Phi({\Lha}) = \frac{\psi\ t_i A_0}{\eta(\beta-\frac{1}{\eta}-1)}\ 
	{\Lha}^{-\beta}\
	\Biggl[1 - \biggl(\frac{\Lha}{{\Lha}_{\rm 0, up}}\biggr)^{\beta-
	\frac{1}{\eta}-1}\Biggl]\ +\ \psi t_i A_0\ {\Lha}^{-\beta}\quad .
\end{equation}
This again implies that the slope of
$\Phi({\Lha})$ remains unaffected by ageing for $\beta >
\frac{1}{\eta}+1$, whereas if $\beta < \frac{1}{\eta}+1$, ageing
produces an observed slope $a=1 + \frac{1}{\eta}$, independent of
$\beta $ and $m_{\rm up}$.  This minimum value of $a$ reflects our
argument above, that the observed slope of $\Phi({\Lha})$ is
steepened if the nebular luminosity evolution is
strong enough.  Figure~\ref{F.lh95.phi} demonstrates this behavior for
$\eta= 5$ and a range of $\beta$, with the curves again normalized to
the same value of ${\Lha}_{0,\rm up}$.  Thus if the slope of
$\Phi({\Lha}_0)$ is 
steep, intrinsically luminous systems are rare and therefore the
faded remnants of such objects make a relatively small contribution to
any bin of observed ${\Lha}$.  Note that this steepening behavior in $a$
was not apparent, for plausible values of $m_{\rm up}$, in the case
where $f(t)$ was set to 0 at time $t=t_{20}$
(equation~\ref{cathie.f}).  In that case, the dynamic range of the
fading function $f(t)$ is small over most of the lifetime of an \hii\
region.  The 
stronger evolutionary effect resulting from equation~\ref{lh95.phi}
stems from the prolonged fading at late times in this prescription.

Note that $\eta$ essentially corresponds to $1/d$ in the previous
case.  Recalling that $d$ is the power law slope for the $m-t_{\rm
ms}$ relation (equation~\ref{mtms}), a value of $\eta=5.0$ suggests $d =
0.2$.  This is considerably shallower than our previous estimate of
0.65, which was fitted from the model results of the Geneva group
(\eg Maeder 1990), the same stellar models used by Leitherer \&
Heckman (1995).  This would suggest that the luminosity evolution of 
the \hii\ regions is affected by additional factors besides the
turnoff of stars from the main-sequence, for example, spectral
evolution of the stars while they are still in the H-burning phase.
Clearer understanding of the behavior of $f(t)$ is needed to confirm
the applicability of our analysis.

\subsection{Discussion of Evolutionary Effects}

We have shown how in principle ageing can cause the measured \hii\ LF
to be steeper than the initial, as cluster fading shifts an
excess of objects into low luminosity bins.  This effect is only
manifest for flatter initial $\Phi({\Lha}_0)$ with low $\beta$, where
any bin of objects at given current luminosity ${\Lha}$
contains an important contribution from initially luminous objects formed
in the distant past.  Thus for that case, it is the fading function
$f(t)$ that determines the observed \hii\ LF.  Conversely, if the
initial $\Phi({\Lha}_0)$ is steep, then any bin of current luminosity 
is dominated by recently formed objects.  In this case, ageing is
unimportant, and the observed $\Phi({\Lha})$ has the same slope as the
initial.  Quantitatively, if the nebula fades according to
a simple power law $t^{-\eta}$, then
\begin{equation}\label{a}
a = {\rm max}\Bigl(\beta,\ a_{\rm min}\Bigr) , \quad 
	a_{\rm min} = 1 + \frac{1}{\eta} \quad .
\end{equation}
Thus for example, if $\eta= 1$, then measured values of $a= 2$
would imply only that $\beta \leq 2$.
Hence, for any simple power-law $f(t)$ that does not
steepen substantially at late times, there is a minimum value
$a_{\rm min}$, independent of $\beta$, given by equation~\ref{a}.
Therefore, if different galaxies have populations of \hii\ regions 
characterised by different values of $\beta$, then this variety would
be manifest only down to a limiting value set by the $f(t)$.  
An observed value of $a_{\rm min} = \frac{1}{\eta}+1$ can therefore be
used to determine the actual slope $\eta$ of the fading function.  For
example, $a_{\rm min} = 1.4$ would imply an
adjustment in the value of $\eta$ from 5 to 2.5, which would require
a much more gradual decline in \Ha\ luminosity for the \hii\ regions
than is predicted by the stellar models.
However, $\eta$ is extremely sensitive to the adopted value of $a_{\rm
min}$, so this test is unlikely to be practical in the near future.
Furthermore, our ability
to observe $a_{\rm min}$ depends on the actual existence of galaxies with
\hii\ LFs having initial $\beta <a_{\rm min}$.  
At present the available data on \hii\ LFs in different galaxies
suggests $a \ga 1.3 - 1.4$ (\eg Banfi {\etal}1993), but there are
not yet enough data to evaluate whether there is a lower cutoff in
the values of $a$.  This is evidently an effect that can be sought in
the future as \hii\ LFs become available for a larger sample of galaxies.

We find, however, that this evolutionary effect of a minimum $a$ is
unlikely to be 
significant for observed \hii\ LFs with spectral indices of $a\sim 2$,
if we use $f(t)$ based on plausible population synthesis arguments.  We have
experimented with two prescriptions:  one assuming
an instantaneous cutoff in ${\Lha}$ at $t>t_{20}$; and
the other using continuous fading, $f(t)\propto t^{-\eta}$, based on
population synthesis models by Leitherer and Heckman
(1995).  In each case, the power-law fading is too steep for the
measured values of $a$ to be explicable as an ageing 
effect, and thus we would conclude that $a$ is a direct measure of
$\beta$.  We caution, however, that this result does depend on the
assumed IMF and modeled stellar parameters that are used in the population
synthesis models.  Some of these parameters, for example, stellar
ionizing fluxes (especially those contributed by B stars), are rather
poorly known.  Furthermore, this determination of $\beta$ from $a$ is
based on stars 
with $m \ga 20\msol$, which is the population ionizing the \hii\
regions.  We are thereby assuming that the slope $\beta$ 
extends to lower-mass stars, which dominate the MLF through their
SNe.  This is based on assumptions about the form of the IMF, and also the
constancy of SN power for stars of different mass.

\subsection{Unsaturated clusters}

In the limit of small cluster membership number $N_*$, there is a
wide dispersion in average \Ha\ contribution per star, according to the
precise stellar membership of the cluster.  As a result, a power law
slope in the IMF produces a $\Phi({\Lha}_0)$ of the same initial slope
at high ${\Lha}_0$, but which flattens at low ${\Lha}_0$ (McKee \&
Williams 1996).  This is produced by the behavior of the scatter,
due to small number statistics, in contributions to ${\Lha}_0$ from a
bin of given $N_*$.  These contributions scatter symmetrically on a linear 
scale of ${\Lha}_0$, but logarithmically, they have a larger displacement
to smaller values of $\log({\Lha}_0)$, producing
a consequent flattening of the luminosity spectrum.  Monte Carlo
simulations by McKee \& Williams (1996) confirm that this flattening occurs
at ${\Lha}_0\sim{\Lha}_{m_{\rm up}}$, the \Ha\ luminosity due to a
single star of $m = m_{\rm up}$.  Clusters with
${\Lha}_0<{\Lha}_{m_{\rm up}}$ are not rich enough to saturate the IMF
up to this maximum, and thus have varying mean \Ha\ contribution per
star; whereas  
those with ${\Lha}_0>{\Lha}_{m_{\rm up}}$, again for a constant IMF, will
have ${\Lha}\propto N_*$, and thereby a constant \Ha\ contribution
per star.  
Note that we do not expect a similar turnover in the
MLF because the relative contribution to $L$ of each star is
dominated by the SN power, which we assume to be independent of $m$.  
The correspondence between $a$ and $\beta$ therefore applies in the
regime where both ${\Lha}_0$ and $L$ scale with the number of stars in
the associations. 

For some galaxies, the slope of the
\hii\ LF shows a turnover to a shallower value at lower ${\Lha}$
(\eg Kennicutt {\etal}1989), which could possibly be due to this
effect, with objects falling below cluster saturation.  The value of
${\Lha}_{m_{\rm up}}$, corresponding to the saturation turnover, is 
currently estimated at $\log {\Lha}_{m_{\rm up}}\sim 38.0 - 38.5$ (Vacca 
{\etal}1996; Panagia 1973).  If the observed flattening of the \hii\
LFs is caused by this effect, it is important to use the slope
$a$ derived from the upper end of the \hii\ LF in estimating
$\beta$.  However, although the observed value of $a$ might 
cause an underestimate of $\beta$ when fitted over the entire range of
${\Lha}$, we will see below that, where the measured $N(R)$
is discrepant from that predicted, it requires {\it shallower} slopes
in the MLF than those inferred from the \hii\ LF.

\section{Comparison with Observations}

We can now compare the prediction with observations for
individual galaxies that have been mapped in \hi\ at sufficient
resolution, and that have measured \hii\ LFs.  Table~\ref{observed}
lists observed parameters for nearby galaxies that we have examined.
Values for the observed slope $a$ of the \hii\ LF,
fitted from data in the listed reference, are
given in Table~\ref{observed}, along with the total number of detected \hii\ 
regions $N_{\HII}$ having ${\Lha}>1\times10^{37}\ergs$, and total
number of detected \hi\ holes $N_{\HI}$.  The limiting 
${\Lha}$ is that at which the surveys are complete for all the galaxies.
We assume that all the \hi\ holes correspond to superbubbles. 
Table~\ref{observed} also lists the assumed \hi\ scale height $h$, 
and reference for the \hi\ mapping.

\def\Hii{\hbox{H{\thinspace}II}}
\def\Hi{\hbox{H{\thinspace}I}}

\begin{table*}
\begin{minipage}{80mm}
\caption{Observed Parameters \label{observed}}
\begin{tabular}{@{}lcrcrcc}
Galaxy & $a$ &  $N_{\HII}$ & \Hii\ Ref.$^{\rm a}$ & 
	 $N_{\HI}$ & $h$(pc) &  \Hi\ Ref.$^{\rm a}$ \smallskip\\

SMC & 1.9 & 93 & (1) & 501 & (2000)$^{\rm b}$ & (3) \\
Holm II & 1.4 & 67 & (2) & 51 & 625 & (4) \\
M31 & 2.1 & 207 & (1) & 140 & 120 & (5) \\
M33 & 2.0 &  257 & (1) & 148 & (100)$^{\rm b}$ & (6) \\
\end{tabular}
\medskip

$^{\rm a}$References: \\
(1)  Kennicutt, Edgar, \& Hodge (1989) \\
(2)  Hodge, Strobel, \& Kennicutt (1994) \\
(3)  Staveley-Smith {\etal}(1996) \\
(4)  Puche {\etal}(1992) \\
(5)  Brinks \& Bajaja (1986) \\
(6)  Deul \& den Hartog (1990) \smallskip\\
$^{\rm b}$Value for the SMC is based on Staveley-Smith's {\etal}(1996)
estimate for the SMC \Hi\ morphology; value for M33 is assumed, not
measured.\\ 

\end{minipage}
\end{table*}

\subsection{Slope of $N(R)$}

Observed slopes for the \hii\ LFs typically fall in the range $1.5
\la a \la 2.5$ (\eg Kennicutt {\etal}1989),
and thus imply a similar range in $\beta$.  We
therefore see that we are indeed in the regime $\beta > \frac{2}{3}$
for which we developed the solutions of $N(R)$ above.  Focusing again
on the superbubble size distribution for $R<R_e$
(equation~\ref{total<}), we find that 
such values of $\beta$ imply an $N(R)$ that is dominated by the
term with a dependence of $R^{1-2\beta}$.  This term describes
stalled shells, including those that are in the survival phase after
stalling.  Our assumed range of $\beta$ therefore implies a range in
slope $\alpha$, where $N(R)\propto 
R^{-\alpha}$, of $2\la\alpha\la 4$.

\begin{figure*}
\begin{center}
   \epsfbox[50 100 500 720]{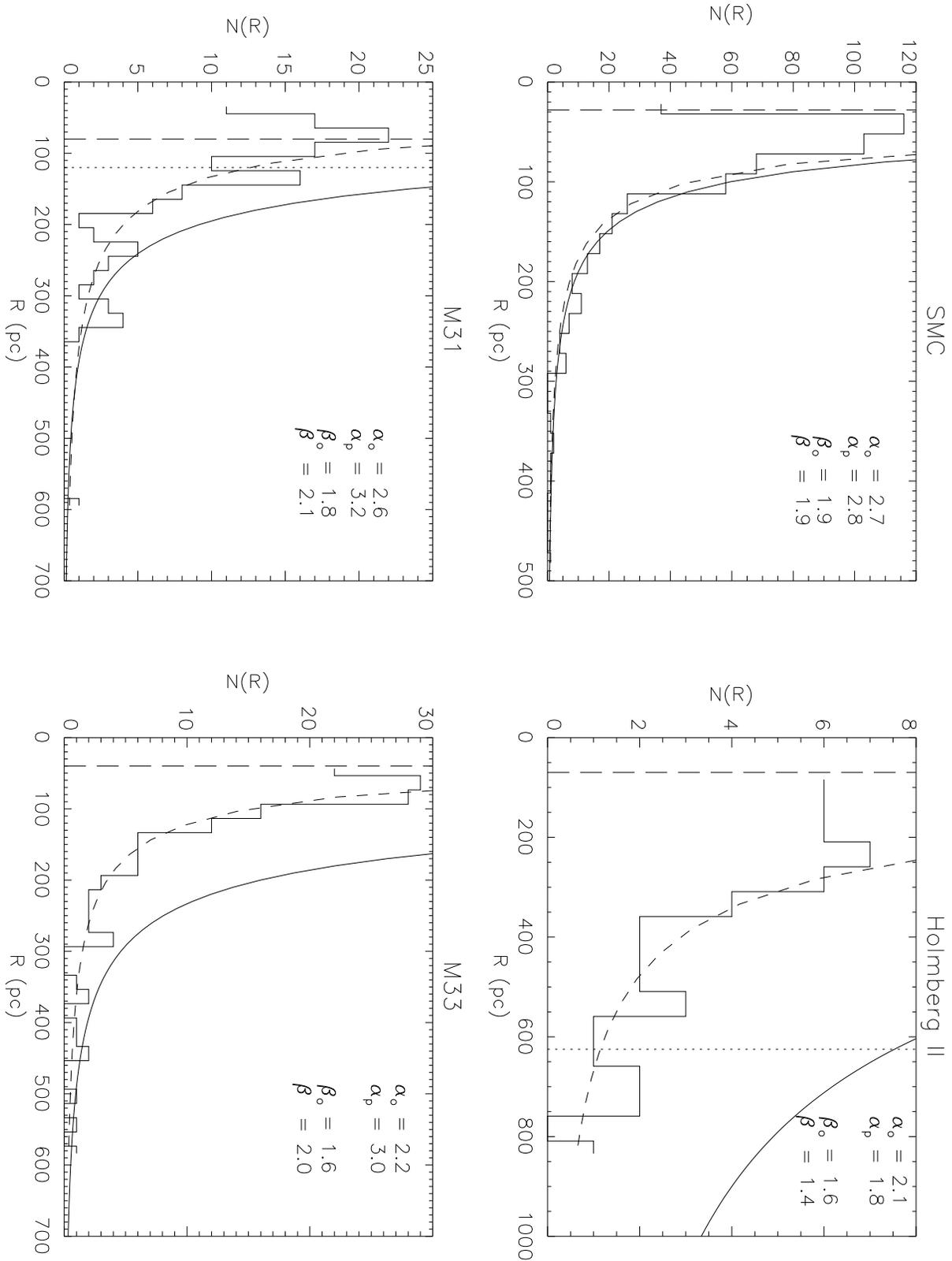}
\end{center}
\caption[]{Histogram of \hi\ hole radii.  Power-law fits to the data
with slope $\alpha_{\rm o}$ are shown with the dashed line, from which 
inferred values of the MLF slope $\beta_{\rm o}$ are computed.
The solid lines show the predicted size distribution with slope
$\alpha_{\rm p}$, computed with the value of $\beta$ from the \hii\ LFs.  
\hi\ scale heights, where known, are indicated by the vertical dotted
lines, and the survey resolution limits are shown by the vertical
long-dashed lines.
\label{linear}}
\end{figure*}

\begin{figure*}
\begin{center}
   \epsfbox[50 100 500 720]{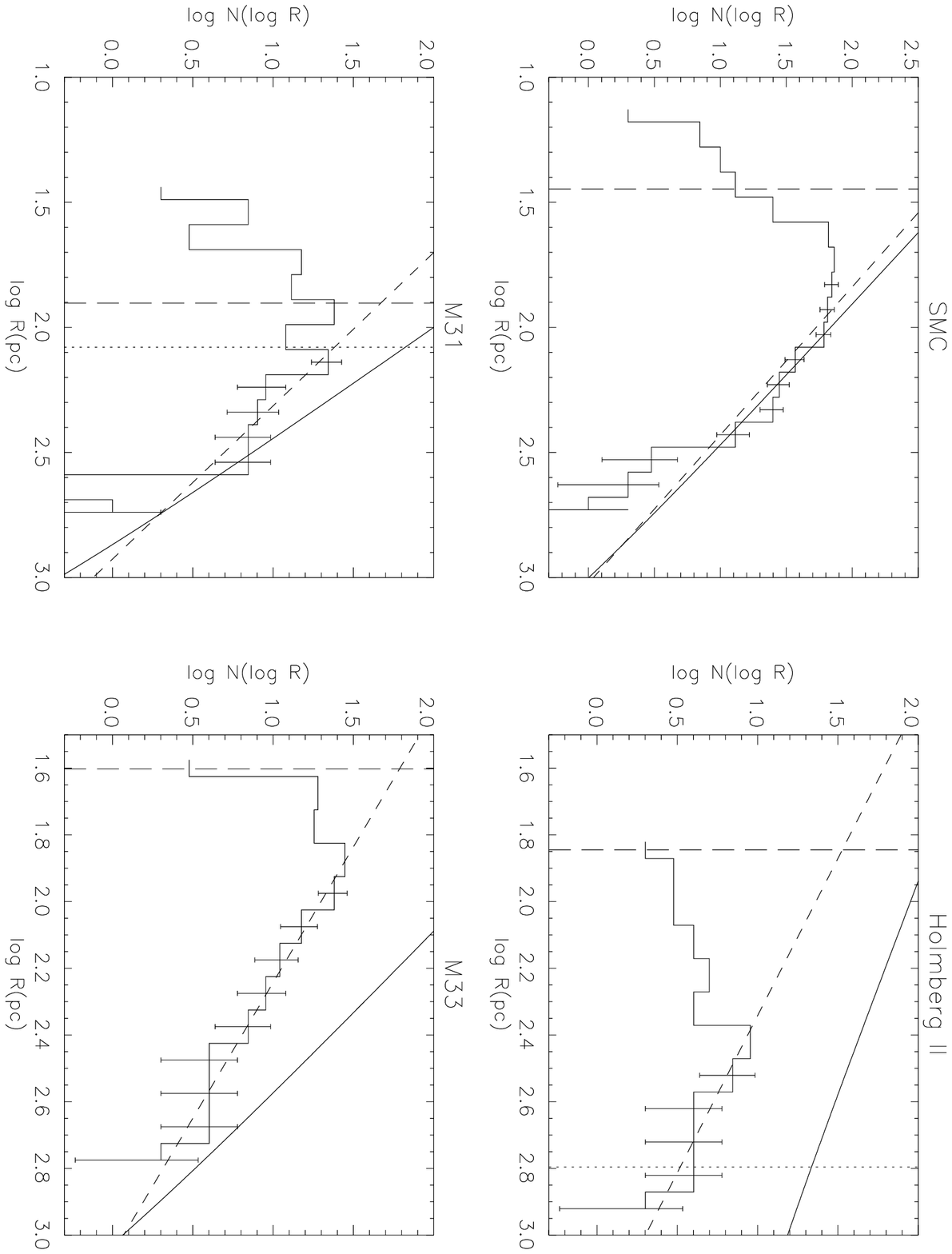}
\end{center}
\caption{Same as Figure~\ref{linear}, displayed and binned on a
logarithmic scale.  Data bins without error bars were not included in
deriving the fitted relation (short-dashed line).  [Note that the
slope of $\log N(\log R)$ vs. $\log R$ is $1-\alpha_{\rm o}$.]
\label{loglog}}
\end{figure*}

Figures~\ref{linear} and \ref{loglog} show
the histograms of \hi\ hole radii in pc, with Figure~\ref{linear}
binned and displayed linearly, and Figure~\ref{loglog} binned and
displayed logarithmically.  The data in the figures are superimposed
with a dashed line showing a least-squares, power-law fit to the data,
of slope $\alpha_o$.  This fit was derived from the logarithmic
binning in Figure~\ref{loglog}, weighted by the inverse of the
root-$N$ errors shown in the figure.  Bins without error bars are
deemed incomplete and were not included in the fit.  We also show the
prediction of equation~\ref{total<} in the solid line, having a slope
of essentially $\alpha_p = 1-2\beta$.  Where known, the \hi\ scale
height $h$ is indicated with the vertical dotted line, and the \hi\
survey resolution limit is indicated by the vertical dashed line.
Values for $\alpha_p$ are
given in Table~\ref{predicted}, and for convenient comparison,
$\alpha_o$ is listed there as well.  We also give values for $\beta_o
= (\alpha_o + 1)/2$, predicted from the observed $\alpha_o$, and the
uncertainties $\sigma(\beta_o)$ derived from the formal standard
deviation on the fit to $\alpha_o$.

\begin{table*}
\begin{minipage}{100mm}
\caption{Inferred Parameters \label{predicted}}
\begin{tabular}{@{}lcccccccc}

Galaxy & $N_{\rm tot}$ & $\psi$(Myr$^{-1}$) & $\log A$ & 
$\beta\thinspace ^{\rm a}$ & $\beta_{\rm o}\thinspace ^{\rm b}$ &
$\sigma(\beta_{\rm o})$ & $\alpha_{\rm p}\thinspace ^{\rm c}$ & 
$\alpha_{\rm o}$ \smallskip\\
SMC & $1.9\times 10^3$ & ~43    &  32.26 & 1.9 & 1.9 & 0.3 & 2.8 & 2.7  \\
Holm II & $1.4\times 10^3$ & ~31 &  13.98 & 1.4 & 1.6 & 1.2 & 1.8 & 2.1 \\
M31 & $4.3\times 10^3$ & ~96  &  39.53 & 2.1 & 1.8 & 0.7 & 3.2 & 2.6 \\
M33 & $5.4\times 10^3$ & 120      &  35.90 & 2.0 & 1.6 & 0.4 & 3.0 & 2.2 \\
\end{tabular}
\medskip

$^{\rm a}$Uncertainties $\beta\pm 0.2$. \\
$^{\rm b}\thinspace\beta_{\rm o} = (\alpha_{\rm o}+1)/2.$ \\
$^{\rm c}\thinspace\alpha_{\rm p} = -1+2\beta.$ \\
\end{minipage}
\end{table*}

The comparison between predicted and observed slopes
is surprisingly good, and is in agreement within the errors for all the
galaxies.  The SMC has by far the most reliable and complete \hi\
data, and shows $\alpha_p = 2.8\pm 0.4$, in excellent agreement
with $\alpha_o = 2.7\pm 0.6$.  This is more vividly demonstrated by the
comparison that $\beta = 1.9 \pm 0.2$, taken from the \hii\ LF, which
is in superb agreement with $\beta_o = 1.9 \pm 0.3$, predicted by the
\hi\ hole size distribution.  The agreement for the SMC is
extremely encouraging, given the high level of completeness in the data
for this galaxy (see below).  
While the agreements for Holmberg~II, M31, and M33 are additionally
gratifying, unfortunately the poor statistics in the data for those
galaxies render the comparison less significant.

\subsection{Normalization of $N(R)$}

The normalization of $N(R)$ depends on the coefficient $A$ of the MLF
and the superbubble formation rate $\psi$.  $A$ can be calculated by
integrating equation~\ref{MLF} over the relevant range of $L$.  This yields,
\begin{equation}\label{A}
A = \bigl(1-\beta\bigr)\ 
	\biggl[L_e^{1-\beta} - L_{\rm min}^{1-\beta}\biggr]^{-1} \quad .
\end{equation}
The lower limit of integration, $L_{\rm min}$, is the power associated
with a single SN explosion.  In the standard treatment for the SN power used
above, this implies $L_{\rm min} = E_{51}/t_e = 8\times 10^{35}\
\ergs$,  where $E_{51} = 10^{51}$ erg, the assumed energy of a single SN
explosion.  In this limit, the approximation breaks down in its treatment
of an energy input over the period $t_e$, and we will address this point
in \S 5.3 below.  However, this does provide a useful value for $L_{\rm min}$.
We use $L_e$ as the upper limit of integration because we are comparing
the observations to equation~\ref{total<}, which applies only for
$R<R_e$, or equivalently, $L<L_e$.  Since $N(R)$ is dominated by
stalled objects at small $R$, and therefore small $L$,
equation~\ref{A} is not sensitive to the choice of upper limit.


The superbubble formation rate $\psi= N_{\rm tot} / (t_e+t_s)$,
where $N_{\rm tot}$ is the total number of superbubbles at any given
time.  This assumes $\psi$ to be constant over the timescale of $t_e=40$ Myr,
relevant to the observations.  As can be seen in
Table~\ref{observed}, it generally appears that for the galaxies
examined, $N_{\HII}>N_{\HI}$.  This is presumably a selection effect
due to the sensitivity and resolution limits of the \hi\ surveys.  
The incompleteness is also better understood for $N_{\HII}$
than $N_{\HI}$, so we will therefore estimate $N_{\rm tot}$ from
$N_{\HII}$.  Since ${\Lha}$ for the nebulae diminishes to below the
detection limit well before $t_e$, we must apply a correction to
$N_{\HII}$ to obtain the total number of superbubbles $N_{\rm tot}$,
which we estimate as follows.

Unsurprisingly, the proximity of the SMC, surveyed at high sensitivity 
in \hi\ (Staveley-Smith {\etal}1996), reveals a much greater surface
density of superbubbles than the other galaxies.
$N_{\HI}>N_{\HII}$ for the SMC, which, as mentioned above, would be
expected generally.  Furthermore, the excellent agreement in observed
and predicted slope for $N(R)$ suggests that for the SMC, we can
simply normalize the predicted $N(R)$ to the data, obtain $N_{\rm
tot}$, and thereby estimate $N_{\rm tot}/N_{\HII}$.  Assuming that
the shell census for the SMC is complete for $R>100$ pc, we can use the
integral $\int_{100\rm pc}^{R_e}N(R)\ dR$
to normalize the prediction for $N(R)$ and obtain $N_{\rm tot}$.  Integrating
equation~\ref{total<} yields,
\begin{equation}\label{Ntot}
N_{\rm tot} = \frac{N_{100}}{2A}\frac{R_e^{2-2\beta}}{L_e^{1-\beta}}
	\bigl(2-2\beta\bigr) \biggl[R_e^{2-2\beta} -
	R_{100}^{2-2\beta}\biggr]^{-1} \quad ,
\end{equation}
where $N_{100} = 159$ is the number of SMC shells with
$R> R_{100}$, which is 100 pc.  Using $\beta = 1.9$ from the \hii\
LF for the SMC (Tables~\ref{observed} and \ref{predicted}), this yields
$N_{\rm tot} = 1.95\times 10^3$, and $\psi = 43\ \rm Myr^{-1}$ with
$t_e + t_s = 45$ Myr.  We therefore obtain $N_{\rm tot}/N_{\HII} = 21.0$. 

General application of this conversion factor
does assume that $N_{\HI}$ is essentially complete at
$R>R_{100}$, although Staveley-Smith {\etal}(1996) believe there is
still substantial incompleteness in their data.  However, based on the
following, we expect that most of the incompleteness is in the small
shells, where the resolution limit of 28 pc becomes important.
The SMC ratio of $N_{\rm tot}/N_{\HII}= 21.0$ implies a typical \hii\ region
age of 2.1 Myr, for $t_e +t_s = 45$ Myr.  We note that if \hii\
regions fade according to the power law 
prescription (equation~\ref{lh95.f}), the average age of nebulae that
are observed before fading below a given completeness limit $L_{\rm
c}$ is given by, 
\begin{equation}
\bar{t} = \frac{\int_{L_{\rm c}}^\infty \phi(L)\ t_{\rm c}(L)\ dL}
	{\int_{L_{\rm c}}^\infty \phi(L)\ dL} \quad ,
\end{equation}
where $t_{\rm c} = t_i\ \Bigl(\frac{L}{L_{\rm c}}\Bigr)^{-1/\eta},$ so that
\begin{equation}\label{meanage}
\bar{t} = \Biggl(\frac{\beta-1}{\beta-1-\frac{1}{\eta}}\Biggr)\ t_i
	\quad .
\end{equation}
The mean age $\bar{t}$ is therefore close to $t_i$ because,
with the assumed luminosity evolution that fades strongly after
$t_i$, most of the nebulae will spend only a short fraction of their lifetimes
in the fading period before disappearing below the completeness limit.
For $\beta=2$, and taking
$t_{\rm ms}(m_{\rm up})$ to be $t_i = 2.8$ Myr (\eg Maeder 1990),
equation~\ref{meanage} gives $\bar{t} = 2.2$ Myr.  Thus the
mean \hii\ region age based on our assumptions for timescales and
luminosity evolution is in excellent agreement with that of 2.1 Myr
implied by the SMC data.  We are therefore fairly confident that
this ratio of $N_{\rm tot}/N_{\HII}$ may be used as a reasonable
conversion factor to obtain $N_{\rm tot}$.  However, we
caution that applying the SMC conversion universally assumes the
same relative extinction for nebulae in all the galaxies, which
is actually likely to vary.  The SMC itself appears to have a large
column depth of \hi\ (Staveley-Smith {\etal}1996), and may have a
disproportionate fraction of \hii\ regions lost below the completeness
limit of the nebular census.  

Using $N_{\rm tot}/N_{\HII}=21.0$, we estimate $N_{\rm tot}$ for the
remaining galaxies.  These are listed in Table~\ref{predicted}, along
with the resultant $\psi$ and MLF parameters $A$, $\beta$.  These
parameters can now be 
used in equation~\ref{total<} to quantitatively predict the superbubble size
distributions for each galaxy, that are shown in Figures~\ref{linear}
and \ref{loglog}.

The relative normalizations of the observations and predictions are in
worse agreement than the slopes, reflecting the huge incompleteness in
the \hi\ hole data described above.  There is also significant
uncertainty in the conversion $N_{\rm tot}/N_{\HII}$, which as we
argued, may be slightly higher in the SMC than the other galaxies.
Hence the disparity in the total numbers of predicted and observed
$N(R)$ may be slightly exaggerated in these other galaxies.  

One check on the critical parameter $\psi$ and resulting
normalizations is a calculation of the expected 
core-collapse (Type II + Ib) SN rate, $\psi_{\rm SN} = \bar N_{\rm
SN}\psi$, that is implied by our analysis.  The average number of SNe per 
cluster $\bar N_{\rm SN}$, can be estimated as 
\begin{equation}
\bar N_{\rm SN} = \frac{\bar L t_e}{E_{51}} = 
	\frac{\bar L}{L_{\rm min}} \quad .
\end{equation}
This again assumes that the stellar winds contribute little to
$L$.  The mean $\log \bar L$ is set by $\beta$ for constant
$L_{\rm min}$.  We tabulate $\log \bar L$ and $\psi_{\rm SN}$ for
the galaxies in Table~\ref{SNrates}, with
SN rates given in SNu = number of SNe per $10^{10}\ L_\odot(B)$
per 100 yr.  Table~\ref{SNrates} also shows for comparison the
empirically estimated core-collapse SN rates $\psi_{\rm SN}$(obs) as
reviewed by van den Bergh \& Tammann (1991).  There is a factor of
a few uncertainty in these rates, but reassuringly, our computed $\psi_{\rm
SN}(\beta)$ are in reasonable agreement or less than the empirical estimates.
Our normalizations for $N(R)$ are therefore generally quite
consistent with the expected SN rates, and at least are not
substantial overestimates.  

\begin{table*}
\begin{minipage}{80mm}
\caption{Supernova Rates \label{SNrates}}
\begin{tabular}{@{}@{}lcclc}

Galaxy & $M_B$(Ref.)$^{\rm a}$ & $\bar L(\beta)$ &$\psi_{\rm SN}(\beta)$ &
	$\psi_{\rm SN}$(obs)$^{\rm b}$ \\
 &  & $\bigl(\ergs\bigr)$ & (SNu) & (SNu) \smallskip\\
SMC & --15.7 (1) & $8.6\times 10^{36}$ & ~~1.1 & 1.3 -- 3.0 \\
Holm II & --16.6 (2) & $6.3\times 10^{37}$ & ~~2.5 & (2.9) \\
M31 & --21.6 (3) & $4.8\times 10^{36}$ & ~~0.006 & 0.03 \\
M33 & --18.4 (1) & $6.3\times 10^{36}$ & ~~0.19 & 0.4 \\
\end{tabular}
\medskip

$^{\rm a}$References for $M_B$: \\
(1)  van den Bergh \& Tammann (1991)\\
(2)  Puche {\etal}(1992)\\
(3)  Sandage \& Tammann (1987) \smallskip\\
$^{\rm b}$From van den Bergh \& Tammann (1991); value for
	Holmberg~II is that estimated for galaxies of type Sdm -- Im, with
	$H_0 = 75\ \kms\ {\rm Mpc}^{-1}$. \\
\end{minipage}
\end{table*}

\subsection{Peak in $N(R)$}

The observations show a peak in $N(R)$, which is undoubtedly related
to the resolution limit of the \hi\ surveys.  However, it turns out
that the theoretical size distribution peaks at similar radii.
Since $N(R)$ is dominated by the stalled shells, the minimum stalled $R =
R_{\rm min}$ is simply the stall radius corresponding to the minimum
input power $L_{\rm min}$, which corresponds to that of individual
SNe, as shown above.  For $R<R_{\rm min},\ N(R)\propto R^{2/3},$ as
given by equation~\ref{E.grow<.b<} (\S 3.3).  Our value of $L_{\rm min} = 8
\times 10^{35}\ \ergs$ thus yields $R_{\rm min} = 30$ pc 
(equation~\ref{Rf}).  However, the evolution of individual SNRs is
rather different from that of superbubbles with continuous energy
injection.  Our limiting extreme in superbubble evolution yields an
SNR stall radius that is a factor of 2 or 3 less than estimates by
Cioffi \& Shull (1991), who consider more realistic SNR evolution and
slightly different ISM parameters.
Consideration of an ambient magnetic field by Slavin \& Cox (1992) produces
maximum radii of $\sim 50$ pc, 20\% less than unmagnetized evolution.
So it would seem that the peak in the size
distribution of \hi\ holes should fall between 30 and 100 pc.  

The \hi\ surveys seem to suggest a peak of $\la 50$ pc.  The
resolution limits have similar values, so at present it is unclear
whether $N(R)$ continues to increase toward smaller radii, or whether
the peak is actually falling in the expected range.  It is interesting
that for the SMC in particular, which has the highest quality data
and resolution limit of 28 pc, that $N(R)$ still shows no sign of
falling off at low values.
However, our treatment of superbubble evolution assumes a universal
lifetime for all shells of $t_e = 40$ Myr.  Again, this assumption
breaks down for individual SNRs.  The peak in the size distribution
may be affected by their shorter lifetime and shrinkage by the
ambient pressure or magnetic tension (Ferri\`ere {\etal}1991) after
the maximum sizes are attained. 
At any rate, assuming that the peak radius does
actually correspond to the typical endstage radius for individual SNRs,
the observed peak value of $R$ could help constrain SNR
evolution, ISM conditions and/or the typical SN energy in galaxies for
which it is well-determined.

\section{Superbubbles and the ISM}

It is extremely encouraging that the predictions and observations for the
slope of $N(R)$ are in agreement for the \hi\ holes in these galaxies,
especially the 
largely complete sample for the SMC.  Since these superbubbles 
constitute one of the principal forms of structure in the ISM, the
implication is that its large-scale structuring is indeed most likely
determined largely by these OB superbubbles.  Furthermore, a strong
agreement, as might be the case for the SMC, implies that {\it no
additional fundamental process is necessary to explain the creation
and evolution of observed \hi\ holes.}  

The agreement of our relation for $N(R)$ with the SMC data
additionally suggests that most of the simplifying assumptions made in
\S 2.1, such as constant IMF, single-burst star formation in clusters,
uniform ambient medium, etc. are practical in this analysis.  We have
confirmed that the MLF spectrum,  
in particular, is a critical parameter to $N(R)$, and affects the 
resulting ISM structure.  In comparing with Galactic \hi\ observations,
Bregman, Kelson, \& Ashe (1993) found discrepancies in ISM structure
functions modeled from a uniform size distribution of \hi\ holes.
It would be interesting to see whether accounting for a power-law
$N(R)$, as required by the MLF, can
more accurately reproduce the observations. 

We do caution that Staveley-Smith {\etal}(1996) note a strong
coherence in the dynamical ages of the SMC shells, suggesting a short
formation burst.  This is in contradiction with our assumption of
constant $\psi$.  We predicted the form of $N(R)$ for a single burst
in \S 3.2, finding that $N(R)\propto R^{1-2\beta}$ for $R<R_f(t_b)$,
where $t_b$ is the age of the burst.  For $R>R_f(t_b),\ N(R)\propto
R^{4-5\beta}$, corresponding to growing shells, with a factor
$\frac{5}{2}$ discontinuous jump.  Staveley-Smith {\etal}(1996) find $t_b =
5$ Myr for a burst scenario, yielding $R_f = 160$ pc.  The data in
Figure~\ref{linear} extend to twice this radius, with no suggestion of
the predicted discontinuity.  The observations therefore appear to be
inconsistent with this single burst.
Since the constant formation model implies that most shells have stalled, we
suggest that the Staveley-Smith {\etal}(1996) result may stem from 
this majority of stalled objects having spurious expansion velocities
attributed to them.  The stalled shell walls may instead reflect random
ISM velocities.  We plan a followup study of the distribution in 
expansion velocities, which will clarify this issue.

As is apparent in Figure~\ref{loglog}, the two disk galaxies, M31 and
M33, show greater disagreement in predicted and observed slopes of $N(R)$ than
do Holmberg~II and the SMC, which are both Magellanic irregulars.
Owing to the apparent enormous 
incompleteness in the \hi\ hole samples and high uncertainties on the
fit to $N(R)$, we cannot attribute significance
to this possible correlation between galaxy morphology and relative
agreement in slopes.  Nevertheless, the contrast in relative agreement
between the different galaxy types is suggestive and interesting. 
Since the gas distribution and dynamical processes in spiral galaxies
differ from those of dwarf irregulars, it is reasonable to suspect that 
superbubble evolution is likely to differ between these morphological types
of galaxies.  For example, Ferri\`ere (1995) shows that the three-dimensional
evolution of superbubbles in an exponential gas disk 
is quite different from the spatially uniform expansion we have assumed here.
Differential rotation in the disks and radial
effects such as the distribution of gas and star-forming regions are
other examples of factors that are likely to be important in spiral galaxies.

\subsection{The Smallest and Largest Shells}

The regimes where we expect significant disagreement between the prediction and
observations would be at the extremes in radius.  As described above,
the smallest shells should correspond mostly to individual SNRs, whose
evolution is significantly different from what is assumed by
equation~\ref{Rcgs}.  The continuous wind approximation is also likely to
break down for superbubbles created by only a few discrete SNe.
The sizes of such objects may be underestimated since observational
and theoretical evidence suggests that many SNRs strike the superbubble walls,
therefore converting their energy into direct kinetic impulses and
shell X-ray radiation instead of thermal energy in the shell interior
(\eg Oey 1996; Chu \& Mac Low 1990; Franco {\etal}1991).
In addition, we may expect a significant contribution from Type Ia
SNe.  However, the addition of Type Ia SNRs to $N(R)$ would
cause a jump in the peak predicted at $R_{\rm min}$.  There is as yet
no evidence for such a jump, again demonstrating that
$R_{\rm min}$ itself is empirically not yet apparent.  

At large $R$, we expect the \hi\ scale height $h$ to
be a critical factor in the evolution of the superbubbles.  The
growth of the superbubble radius to $h$ allows the interior hot gas
to break out of the galactic disk and depressurize the shells.  Heiles
(1990) assumed a bimodal evolution, where for $R<h$, the shell growth
followed $R\propto t^{3/5}$ given by equation~\ref{Rcgs}; and for $R>h$,
the growth was described by a coasting, momentum-conserving phase with
$R\propto t^{1/3}$ (cylindrical geometry).  Such an effect should cause a 
steepening in slope $\alpha$ of the size distribution at large $R$:  
since these shells can no longer grow as large as they would have done
adiabatically, they accumulate at smaller final radii.  However, the
observations are showing slopes that are generally shallower than $\alpha_p$,
rather than steeper, thus the limiting effect of $h$ is not apparent
in the data for these galaxies.  As seen in Figure~\ref{linear},
$N(R)$ does appear rather irregular 
in M31, especially at $R$ a few tens of pc above $h$, hinting at
the possible effect of breakouts.  However, for M33 and the SMC,
$N(R)$ follows a power-law distribution quite smoothly, even for $R>h$.
There is a hint of a break in the SMC data around 300 pc, that could
possibly correspond to an equivalent $h$, which is indeterminate for
this galaxy.  

It seems likely that some other mechanism might also create superbubbles
with $R>h$, generating an excess of large superbubbles, and thereby
flattening the observed slope $\alpha_o$ of the size distribution.
As discussed by Heiles (1990), among others, the two principal
candidate mechanisms are propagating star formation and infalling
high-velocity clouds (HVCs).  Propagating star formation (\eg McCray
\& Kafatos 1987) produces generations of OB associations in close
spatial proximity, thereby increasing the mechanical power $L$
and extending the duration of energy input beyond $t_e$.  This can
therefore lead to supergiant shells with radii larger than are likely
to be due to individual OB associations.  Impacts by HVCs have also been
demonstrated to produce shell structures with radii of $10^2 - 10^3$ pc
(\eg Tenorio-Tagle {\etal}1987; Rand \& Stone 1996).  But these
mechanisms, if applicable, apparently do not create enough shells to
clearly distinguish their contribution to the superbubble size distribution
in the current data, and blend smoothly with $N(R)$ predicted by the
MLF for OB associations.

It is also quite possible that many of the larger shells are created
by the merging 
of smaller ones.  This would cause a flattening in the slope $\alpha$
of the size distribution, since smaller shells would be eliminated to
combine larger ones.  The inter-cluster distance for OB
associations is typically a few hundred pc in these galaxies, so we
expect merging to be important for shells with radii larger than this
range.  We note that the observed slopes are all
flatter than predicted, with the exception of that for Holmberg~II,
which has by far the worst uncertainty.  We commented above on the
possibility of disagreement between predicted and observed slope in the spiral
galaxies versus the Magellanic irregulars; another potential
explanation for such a discrepancy is that the concentration of
star forming regions in the spiral arms enhances merging and
propagating star formation, preferentially encouraging the production
of the very large shells and thereby flattening the resultant~$\alpha$.  

\subsection{Porosity of the ISM}

The favorable comparison between the observations and prediction
encourage us to apply this analysis to the global structure of the ISM
in galaxies.  A fundamental outstanding issue is the relative
importance of hot, coronal gas, which presumably originates in
superbubbles and Type~Ia SNRs, in relation to the cooler phases of the
ISM.  This question is traditionally addressed by means of
the porosity parameter $Q$ (Cox \& Smith 1974), which is the ratio of
total volume or area occupied by superbubbles to the total volume or
area of the host galaxy.  Following directly on the preceding section,
we can also evaluate the degree of shell merging or overlap by examining 
$Q$.  

Following Heiles (1987, 1990), we compute both the two-dimensional
$Q_{\rm 2D}$, and volume porosity $Q_{\rm 3D}.~~ Q_{\rm 2D}$ is especially
appropriate for the 
disk distribution of OB associations in spiral galaxies.  We have,
\begin{equation}
Q_{\rm 2D} = A_g^{-1}\ \int A_b\ N(A_b)\ dA_b \quad ,
\end{equation}
where $A_b = \pi R^2$ is the projected superbubble area, and $A_g$ is the
area of the galactic disk.  Rewriting this expression in terms of
$N(R)\ dR$:
\begin{equation}
Q_{\rm 2D} = (\pi R_g)^{-2}\ \int_{R_{\rm min}}^{R_e} \pi R^2\ N(R)\ dR \quad ,
\end{equation}
recalling that we are in the regime $R < R_e,\ \beta >
\frac{2}{3}$ for the standard model.  We obtain,
\begin{equation}\label{Q2D}
Q_{\rm 2D} = R_g^{-2}\ A\psi L_e^{1-\beta}\ R_e^2\ 
	\Biggl\{\biggl[\frac{B}{4-2\beta} + \frac{C}{5-2\beta}\biggr] - 
	\biggl(\frac{R_{\rm min}}{R_e}\biggr)^{4-2\beta}
	\biggl[\frac{B}{4-2\beta} + \frac{C}{5-2\beta}
	\biggl(\frac{R_{\rm min}}{R_e}\biggr)\biggr]\Biggr\} \quad ,
\end{equation}
where $R_g$ is the radius of the galactic disk, and 
\begin{equation}
B = 2(t_e + t_s) 
\end{equation}
\begin{equation}
C = \frac{~9-6\beta}{-2+3\beta}\ t_e \quad .
\end{equation}
For $\beta > 2,\ Q_{\rm 2D}$ is dominated by $R_{\rm min}$, but for
$\beta < 2,$ the largest shells dominate.  Interestingly, the
relevant values of $\beta$ fall in this transition.  Thus
for some galaxies, the few largest shells dominate $Q_{\rm 2D}$, whereas
in others it is dominated by the many individual SNRs.  For
those with $\beta\sim 2$, the relative superbubble sizes
contribute fairly equally in determining $Q_{\rm 2D}$.  

The $C$ term, corresponding to the term in equation~\ref{total<} with
dependence $R^{2-2\beta}$, is small for parameters of interest, and
will only dominate for $\beta< 1$.  We may therefore approximate,
\begin{equation}
Q_{\rm 2D} \simeq R_g^{-2}\ A\psi L_e^{1-\beta}\ R_e^2\ 
	\frac{2(t_e+t_s)}{4-2\beta}\ 
	\Biggl[1-\biggl(\frac{R_{\rm min}}{R_e}\biggr)^{4-2\beta}\Biggr]
	\quad .
\end{equation}

By analogy, we also compute the three-dimensional porosity parameter:
\begin{equation}\label{Q3D}
Q_{\rm 3D} = \frac{2}{3hR_g^2}\ A\psi L_e^{1-\beta}\ R_e^3\ 
	\Biggl\{\biggl[\frac{B}{5-2\beta} + \frac{C}{6-2\beta}\biggr] - 
	\biggl(\frac{R_{\rm min}}{R_e}\biggr)^{5-2\beta}
	\biggl[\frac{B}{5-2\beta} + \frac{C}{6-2\beta}
	\biggl(\frac{R_{\rm min}}{R_e}\biggr)\biggr]\Biggr\} \quad ,
\end{equation}
using the \hi\ scale height $h$ as the relevant galactic vertical
extent.  $R_{\rm min}$ dominates for $\beta>2.5$, so
$Q_{\rm 3D}$ is almost always dominated by the largest shells.  Again, we
find that the $C$ term is usually small, dominating only for $\beta
\la 1$, so we may approximate,
\begin{equation}
Q_{\rm 3D} \simeq \frac{2}{3hR_g^2}\ A\psi L_e^{1-\beta}\ R_e^3\ 
	\frac{2(t_e+t_s)}{5-2\beta}\ 
	\Biggl[1-\biggl(\frac{R_{\rm min}}{R_e}\biggr)^{5-2\beta}\Biggr]
	\quad .
\end{equation}

We take $R_g$ to be the distance out to which \hi\ holes are detected 
in these galaxies.  The adopted $R_g$, and resultant $Q_{\rm 2D}$ and
$Q_{\rm 3D}$ are given in Table~\ref{Q}.  These are the exact values,
computed from equations~\ref{Q2D} and \ref{Q3D}.  We also list the
observed values of $Q_{\rm 2D}$ and $Q_{\rm 3D}$, computed from the \hi\ holes
with $R> R_{\rm min}$.  We caution that the predicted porosities could be
overestimates, since equations~\ref{Q2D} and \ref{Q3D}
assume a size distribution extending to $R_e = 1300$ pc.  Because of
the small numbers of large shells predicted, it is unclear whether the
distributions indeed may be considered to extend to $R_e$.

\begin{table*}
\begin{minipage}{70mm}
\caption{Porosity Parameters \label{Q}}
\begin{tabular}{@{}@{}lcllll}
Galaxy & $R_g$ & \multicolumn{2}{c}{Predicted} & 
	\multicolumn{2}{c}{Observed} \\
 & (kpc) & $Q_{\rm 2D}$ & $Q_{\rm 3D}$ & $Q_{\rm 2D}$ & $Q_{\rm 3D}$ 
	\smallskip\\
SMC & ~2 & 2.1 & 0.3 & 1.6 & 0.1 \\
Holm II & ~7 & 1.5 & 1.2 & 0.1 & 0.08 \\
M31 & 17 & 0.03 & 0.06 & 0.01 & 0.02 \\
M33 & ~7 & 0.3 & 0.9 & 0.08 & 0.2 \\
\end{tabular}
\end{minipage}
\end{table*}

For M31 and M33, we take $Q_{\rm 2D}$ to be the relevant porosity
parameter, and for the SMC, $Q_{\rm 3D}$.  The values in Table~\ref{Q} of
the relevant porosity parameters for these galaxies are all
significantly less than 1.  We 
therefore do not expect a great deal of superbubble overlap and
merging in these galaxies, thereby limiting the extent and networking
of the hot gas that originates within the shells.  
On the other hand, the predicted porosity parameters for Holmberg~II are
$\ga 1$.  This galaxy has a very large
scale height, so both $Q_{\rm 2D}$ and $Q_{\rm 3D}$ are relevant, as is
indicated by their similar values.  These predict that Holmberg~II,
unlike the other galaxies, does have an ISM that is dominated by
coronal gas.  This supports the expectation that the
importance of the HIM depends on the level of star formation
relative to galaxy size and interstellar conditions.
It is important to keep in mind that, as seen in \S 5.2, 
the normalizations of $N(R)$ are uncertain, and these affect the
derived porosity parameters, as does the assumption of a distribution
extending to $R_e$.  However, given the general consistency
with the SN rates, and possible overestimate of the $N_{\rm
tot}/N_{\HII}$ conversion factor, we imagine that the
porosities in Table~\ref{Q} should not be hugely underestimated.
Our derived $Q_{\rm 2D}$ nevertheless essentially agree with the values
estimated by Heiles (1990) for M31 and M33.  His analysis is based on
somewhat different criteria with greater uncertainties.  For example,
he examined primarily superbubbles that break out of the
galactic disk, and included an {\it ad hoc} adjustment in his
${\Lha}-N_*$ relation.

Note that the existence of the MLF spectrum, which accounts for the
clustering of core-collapse SNe, has a potentially important effect on
the porosity.  Whereas $N_*$ uniformly distributed SNe will, at their
final radii, contribute to 
$Q_{\rm 3D}$ simply as $N_*L_{\rm min}^{3/2}\propto N_*$, the
same SNe concentrated into one cluster will contribute as 
$(N_* L_{\rm min})^{3/2}\propto N_*^{3/2}$ (for $N_* L_{\rm min} <
L_e$; equation~\ref{Rf}).  Clustered SNe in superbubbles therefore
produce a larger $Q_{\rm 3D}$ than the same number of individual SNe.
However, this is not the case for $Q_{\rm 2D}$, where for both individual
and clustered SNe, $Q_{\rm 2D}\propto N_* L_{\rm min}$.  Although, as we have
seen above, the largest shells constitute a larger component of
$Q_{\rm 2D}$, the relative degree of clustering will not affect the 
actual value of $Q_{\rm 2D}$.  Ferri\`ere (1995) also finds a strong
effect of clustering on $Q_{\rm 3D}$, using a more complex model for shell
evolution in an exponential gas disk.  

It is also important to bear in mind the contribution of Type~Ia
SNRs (\eg Slavin \& Cox 1993), although in most star-forming galaxies,
these are outnumbered by factors of 3--10 by core-collapse SNe
(van den Bergh \& Tammann 1991).  
Heiles (1987) emphasizes that the distribution of Type~Ia SNe should vary
with galactocentric radius in disk galaxies, owing to the distribution
of progenitors, and could therefore cause important radial effects in the ISM
porosity.  This effect could be a factor in the discrepancy between
observed and predicted slopes for M31 and M33.  As mentioned
in \S 6.1, the \hi\ data do not yet provide 
empirical evidence for the contribution of Type~Ia SNRs in the
galaxies we have examined.

It is interesting to apply our analysis to the porosity of the Milky
Way.  The \hii\ LF for the Galaxy has been compiled by McKee \&
Williams (1996; hereafter MW96) and Smith \& Kennicutt (1989;
hereafter SK89), where both sets of
authors used primarily the compiled data of Smith, Biermann, \& Mezger
(1978).  The \hii\ LF estimated by MW96 (their
Figure~2), shows a slightly flatter slope of $a=2.0$, compared to
$a=2.3$ reported by SK89.  The data from these papers
yield parameters for the Galaxy shown in Table~\ref{MW}.  
The values of $\beta$ from the \hii\ LF predict a fairly steep slope
for $N(R)$ in the Galaxy of 3.0 and 3.6.
For the SK89 data, we base our calculations on
an estimated total of 145 \hii\ regions having $\log {\Lha}> 37.84\
\ergs$ (see their Figure~1).  It is apparent that the predicted SN
rates $\psi_{\rm SN}$ are an order of magnitude smaller than the
observed $\psi_{\rm SN}$ of $\sim 3$ SNu estimated for the core-collapse
SNe in the Galaxy
(van den Bergh \& Tammann 1991).  We therefore also compute
implied parameters based on the empirical SN rate, for $\beta = 2.0$
(model E2.0) and $\beta=2.3$ (model E2.3).  Resulting
porosities are computed assuming $R_g\sim 13$ kpc, based on the data
of Smith {\etal}(1978), and $h=100$ pc (Kulkarni \& Heiles 1987).

\begin{table*}
\begin{minipage}{120mm}
\caption{Parameters for the Galaxy$^{\rm a}$ \label{MW}}
\begin{tabular}{@{}@{}lccccccccl}
Model & $N_{\rm tot}$ & $\psi$ & $\log A$ & $\beta$ & $\alpha_{\rm p}$ & 
	$\bar{L}(\beta)$ & $\psi_{SN}\thinspace ^{\rm b}$ & $Q_{\rm 2D}$ & 
	$Q_{\rm 3D}$ \\
 & & (Myr$^{-1}$) & & & & $(\ergs)$ & (SNu) & & \smallskip\\

MW96 & $6.5\times 10^3$ & $1.4\times 10^2$  & 35.90 & 2.0 & 3.0 & 
	$6.3\times 10^{36}$ & 0.05 & 0.1 & ~0.3 \\
SK89 & $3.8\times 10^4$ & $8.5\times 10^2$ & 46.78 & 2.3 & 3.6 & 
	$3.1\times 10^{36}$ & 0.14 & 0.2 & ~0.4 \\
\\
E2.0 & $4.0\times 10^5$ & $8.9\times 10^3$ & 35.90 & 2.0 & 3.0 &
	$6.3\times 10^{36}$ & (3.0) & 6.7 & 19 \\
E2.3 & $8.0\times 10^5$ & $1.8\times 10^4$ & 46.78 & 2.3 & 3.6 &
	$3.1\times 10^{36}$ & (3.0) & 3.9 & ~8.1 \\
\end{tabular}
\medskip

$^{\rm a}$Assuming $R_g = 13$ kpc, $h = 100$ pc, and
	$L_{\odot,B} = 2.3\times 10^{10}$. 
\smallskip\\
$^{\rm b}$Value for models E2.0 and E2.3 is the empirical
	estimate of $\psi_{\rm SN}$ for the Galaxy (van~den~Bergh
	\& Tammann 1991). \\
\end{minipage}
\end{table*}

Unfortunately, this discrepancy in the predicted and observed SN rates
has a critical impact on the Galactic porosity parameters.  As seen
in Table~\ref{MW}, the porosity estimates based on the observed
Galactic \hii\ LF are consistent with the low porosities estimated for
the other galaxies (Table~\ref{Q}).  However, the Galactic estimates
based on the much larger, observed $\psi_{\rm SN}$ produce porosities
$\gg 1$, implying a strong dominance of the hot interstellar component.
It is unclear how this discrepancy should be reconciled.  On the one
hand, the observed Galactic SN rate is consistent with the presumed
Hubble type around Sb (van den Bergh \& Tammann 1991).  On the other
hand, van den Bergh finds a similar discrepancy when predicting
$\psi_{\rm SN}$ from the local IMF, which also underestimates the
observed $\psi_{\rm
SN}$ by an order of magnitude (van den Bergh \& Tammann 1991).  We
note that Tammann does not find such a discrepancy, and also that MW96
do find consistency between their \hii\ LF and the Galactic SN rate.
It is interesting to note that the only other Sb galaxy in our study,
M31, shows a similar, but smaller, discrepancy in observed and
predicted $\psi_{\rm SN}$.  

Other authors have estimated the porosity of the Galaxy based on
similar analyses.  Ferri\`ere (1995) finds $Q_{\rm 3D} \sim 0.2$ based on
a more complex, 3D model for shell evolution in an exponential gas
disk, including an important shell contraction phase.
Slavin \& Cox (1993) estimate a value of $Q_{\rm 3D}=0.18$ for individual SNRs,
but exclude the contribution of larger superbubbles, which we found
above to dominate the porosity in our analysis.  Heiles (1990)
estimates $Q_{\rm 2D}=0.30$, ignoring the contribution of shells that do
not break out of the Galactic disk.  Since the measurements of
$\beta$ for the Galaxy fall in the regime where small shells dominate
$Q_{\rm 2D}$, this value is also likely to be a substantial
underestimate.  Our results seem very broadly consistent with
previous studies, but highlight the uncertainties in
Galactic parameters.  Apparently we have simply restated the problem
of whether or not the hot component of the ISM should strongly dominate in
the Milky Way, since this argument is based on the observed supernova
rate in the Galaxy (Cox \& Smith 1974; McKee \& Ostriker 1977).

\section{Conclusion}

We have used the standard, adiabatic shell evolution to predict the
differential size distribution $N(R)$ for populations of OB
superbubbles in a uniform ISM.  The results are strongly
dependent on the inclusion of a power-law MLF for the OB associations.
Another fundamental ingredient is the criterion that the shell growth
stalls upon pressure equilibrium with the environment.  This
condition, along with the given characteristic time $t_e$, 
determines the characteristic radius $R_e$, that divides the
superbubble population into two regimes of solutions corresponding to
$N(R<R_e)$ and $N(R>R_e)$.  $R_e$ and the corresponding $L_e$
therefore make convenient defining criteria for a 
different scale phenomenon, for example, a starburst event.
For the condition of constant shell creation and power-law MLF, 
in the regime $R<R_e,\ N(R)$ is given by equation~\ref{total<}, which
has contributions from growing, stalled, and surviving objects.  For
reasonable values of the MLF slope $\beta \sim 2\pm 0.5$, the size
distribution is dominated by stalled objects, yielding $N(R)\propto
R^{1-2\beta}$.  This $R^{1-2\beta}$ dependence appears to be 
fairly robust, since it applies to the standard, momentum-conserving
evolution (Steigman {\etal}1975) as well.  It also describes $N(R)$
for the single-burst creation of shells, for $R$ smaller than 
the stall radius associated with the burst age.
On the other hand, in the regime $R>R_e$, we find the
dependence $N(R)\propto R^{4-5\beta}$ (equation~\ref{total>}), a 
much steeper relation essentially composed of growing objects.

To estimate $\beta$, we adopt the observed slope $a$ of the \hii\ LF.
However, the observed $a$ could in principle be steeper than the initial 
$\beta$ depending on the relation between the luminosity fading of
the \hii\ regions and the initial slope $\beta$.  We investigated this
problem for power-law nebular fading, and found the existence of a
minimum slope $a_{\rm min}=1 + \frac{1}{\eta}$ (equation~\ref{a}), which is 
determined by the power-law index $\eta$ of the fading function.  Thus, for
$\beta>a_{\rm min}$, we may indeed adopt $\beta = a$, but for
$\beta<a_{\rm min},\ a$ provides only an upper limit to $\beta$, as
the observed slope is indeed steepened to the value $a_{\rm min}$.
Observed slopes of galactic \hii\ LFs are all greater than the
predicted $a_{\rm min}$, so these should provide a fairly reliable
measure for the slope 
of the MLF.  It will be interesting to see whether a lower cutoff in the
distribution of observed $a$ in galaxies will become manifest at
$a_{\rm min}$.  We also found that the existence of this minimum
$a_{\rm min}$ depends on a long-term, power-law nebular fading law,
whereas if the \hii\ region luminosities are instantaneously
extinguished at a specified age, no $a_{\rm min}$ will be observed.
We also derive a useful expression for the mean
\hii\ region age of a complete population of nebulae brighter than a
given luminosity.  This is given by equation~\ref{meanage}, in terms of
$\beta$ and $\eta$.

We compared the predicted size distributions with observed \hi\ hole
distributions in four galaxies.  The recent data for the SMC 
(Staveley-Smith {\etal}1996) appear to be largely complete, and show
excellent agreement with our predicted relation of $N(R)\propto
R^{1-2\beta}$.  Despite the fact that our critical assumptions about
\eg the endstage evolution, are highly uncertain and crudely treated, 
the slope of the size distribution in this galaxy can be entirely
explained by our prediction.  {\it No other fundamental processes are
necessary to explain the observed \hi\ hole distribution in the SMC.}
This furthermore suggests that our assumptions of
constant IMF and coeval star formation in OB associations are
broadly useful on global scales.

The observed and predicted slopes of $N(R)$ are also in agreement for
the three other galaxies we examined.  It
is premature to draw conclusions based on these other comparisons, since
the \hi\ hole data are largely incomplete, as evidenced by the
relative numbers of \hii\ regions vs. \hi\ holes.  The predicted SN rates based
on our analysis also support this conclusion, and are in reasonable
agreement with observed rates.  However, it is intriguing that
the spiral galaxies appear to suggest greater disagreement in
the slope of $N(R)$ compared to the
Magellanic irregulars.  We anticipate that radial and disk properties
associated with spiral galaxies should cause differences in the
observed superbubble size distributions.  The effect of the disk
scale height should be especially apparent in these galaxies, limiting
the growth of shells at these sizes, yet such an effect is not
apparent in the data, which show plenty of shells at larger radii.
Processes such as propagating star formation and merging could
contribute large objects that counteract the expected dearth of large
shells.  At present, however, the data for the spiral galaxies,
especially M33, are adequately fit by a simple
power-law distribution, and there is no strong evidence
for a change in characteristics of large superbubbles.  

We also predict a peak in the size distribution corresponding to the
smallest stalled shells, which would be individual SNRs.  The spatial
resolution of the data do not yet allow meaningful comparison with
prediction.  The contribution of Type~Ia SNRs, which could be
substantial (Heiles 1987), should also be apparent near this peak.

Our derivation is easily applied to estimating the porosity of the
ISM.  We find that, not including Type~Ia SNRs, $Q_{\rm 2D}$ is dominated
by individual SNRs for $\beta > 2$, but the few largest
shells for $\beta < 2$.  On the other hand,
$Q_{\rm 3D}$ is usually dominated by the largest superbubbles, rather than the
multitude of small shells and SNRs.  Our estimates for the porosities
in these galaxies generally show values substantially $< 1$, with the
exception of Holmberg~II.  This therefore predicts that this galaxy would
be dominated by a hot interstellar medium, whereas the others would not.
Furthermore, merging would not be expected to be a dominant process in
the galaxies with porosities $\ll 1$, although merging could still 
flatten the slope of $N(R)$ if it is encouraged by spatially clustered shell
distributions.  Our porosity estimates for these external galaxies are 
in good agreement with previous calculations (\eg Heiles 1990), with
the caveat that they are dependent on the normalizations for $N(R)$.  

However, in predicting the SN rate and porosity of the Milky Way, we
find a critical discrepancy between the predicted and observed SN
rates: $\psi_{\rm SN}$ based on the observed \hii\ LF underestimates the
observed value by an order of magnitude.  The predicted value of $\psi_{SN}$
leads to porosities $< 1$, similar to results for the other galaxies.
However, the observed SN rate yields porosities $\gg 1$, implying a
strong dominance of the HIM.  It is unclear how this discrepancy should
be reconciled, since we have simply recast this pre-existing problem for
the Galaxy. 

Our results suggest that OB superbubbles are indeed a dominant source
of structure in the ISM of galaxies.  We used the simplest
and crudest formulations to derive the superbubble size distribution,
with the aim of identifying dominant processes and evolutionary
features.  The agreement of this derived prediction with
the high-quality SMC data is unexpectedly excellent and 
encouraging.  This tentatively suggests that we have 
fairly successfully identified the basic
features of shell evolution, ISM parameters, and stellar
parameters that are relevant to the global structure and evolution of
superbubbles.  
As new, high-resolution \hi\ data become available for
more galaxies (\eg Thilker 1997; Kim {\etal}1997), we will be able to
more rigorously test these results and distinguish unresolved issues.

\section*{Acknowledgments}

We have enjoyed discussions with many people,
both in Cambridge and in the course of two meetings.  It is a pleasure
to thank Eric Blackman, Don Cox, Laurent Drissen, Pepe Franco, 
Despina Hatzidimitriou, Claus Leitherer, Carmelle Robert, John Scalo,
Mike Shull, and Guillermo Tenorio-Tagle.  We are also grateful to
Lister Staveley-Smith for providing access to the SMC \hi\ data in
advance of publication.

\end{document}